 \documentclass{emulateapj}

\usepackage{graphics}      
\usepackage{amsmath}
\usepackage{amssymb}

\def\wig#1{\mathrel{\hbox{\hbox to 0pt{%
          \lower.5ex\hbox{$\sim$}\hss}\raise.4ex\hbox{$#1$}}}}

\def\etal{{et~al.\,}}

\def\mj{M$_{\rm J}\,$}

\def\rj{R$_{\rm J}\ $}

\def\teff{T$_{\rm eff}\,$}

\def\Dwa{$\,$\uppercase\expandafter{\romannumeral5}$\,$}
\def\mic{$\mu$m$\,$}

\def\sles{\lower2pt\hbox{$\buildrel {\scriptstyle <}
   \over {\scriptstyle\sim}$}}
\def\sless{\lower2pt\hbox{$\buildrel {\scriptstyle <}
   \over {\scriptstyle\sim}$}}
\def\sgreat{\lower2pt\hbox{$\buildrel {\scriptstyle >}
   \over {\scriptstyle\sim}$}}
\def\reference{\bibitem}

\begin{document}

\title{Scientific Return of Coronagraphic Exoplanet Imaging and Spectroscopy Using WFIRST}

\author{Adam Burrows\altaffilmark{1}}

\altaffiltext{1}{Department of Astrophysical Sciences,
Peyton Hall, Princeton University, Princeton, NJ 08544; burrows@astro.princeton.edu}

\begin{abstract}

In this study, we explore and review the scientific potential for exoplanet characterization
by a high-contrast optical coronagraph on WFIRST/AFTA.  We suggest that the heterogeneity in albedo spectra and
planet/star flux ratios as a function of orbital distance, planet mass, and
composition expected for the giant exoplanets at separations from their primaries
accessible to WFIRST will provide unique constraints on giant planet atmospheres, evolution, aerosol and cloud 
properties, and general theory.  Such exoplanets are not merely extrapolations of Jupiter and Saturn,
but are likely to occupy a rich continuum of varied behaviors.  Each in themselves and jointly, optical spectra, 
photometry, and polarization measurements of a diverse population of giant exoplanets in the solar neighborhood
has the potential to reveal a multitude of fundamental features of their gas-giant chemistry, 
atmospheres, and formation. Such a campaign will enrich our understanding of this class of planets
beyond what is possible with even a detailed exploration of the giants in our own solar system,
and will compliment ongoing studies of exoplanets in the infrared and on close-in orbits 
inaccessible to coronagraphy. 

\end{abstract}

\section{Introduction}
\label{introduction}

Exoplanet science is observational and
must rely on the astronomical tools of remote spectroscopic sensing to
infer the physical properties of individual planets and their atmospheres.  Therefore, there is a premium on
obtaining {\it spectra} and on improving sensitivity, without the luxury of the direct, in-situ probes employed
so profitably in our solar system.

Before the successful emergence of the RV (radial-velocity) and transit methods, astronomers
expected high-contrast direct imaging, that separated out the light of planet and star
and provided photometric and spectroscopic data for each, would be the
leading means of exoplanet discovery and characterization.  A few wide-separation
brown dwarfs and/or super-Jupiter planets were detected by this means, but the
yield was meager. The fundamental problem is two-fold: 1) the planets are intrinsically dim,
and 2) it is difficult to separate out the light of the planet from under the glare of the star
for planet-star separations like those of the solar system.  Imaging systems need to suppress the
stellar light scattered in the optics that would otherwise swamp the planet's signature.  The planet/star contrast
ratio for Jupiter is $\sim$10$^{-9}$ in the optical and $\sim$10$^{-7}$ in the mid-infrared.
For Earth, the corresponding numbers are $\sim$10$^{-10}$ and $\sim$10$^{-9}$.  These numbers
are age, mass, orbital distance, and star dependent, but demonstrate the challenge. What is more,
contrast capabilities are functions of planet-star angular separation, restricting
the orbital space accessible.

However, things are changing.  Recently, giant exoplanets and brown dwarfs such as HR 8799bcde, 
$\beta$-Pictoris b, GQ Lup b, 2MASS 1207b, and GJ 504b have been discovered and partially characterized using
near-IR high-contrast techniques. The HR 8799 (Marois et al. 2008,2010; Barman et al. 2011)
and $\beta$-Pictoris b (Lagrange et al. 2009) giant planets/brown dwarfs have masses of $\sim$5 $-$ 15 Jupiter
masses (M$_{\rm J}$) and angular separations between $\sim$0.3 and $\sim$1.5 arcseconds.
Their contrast ratios in the near infrared are $\sim$10$^{-4}$, but capabilities near 10$^{-5}$
have been achieved. Such direct imaging is currently most sensitive to wider-separation ($\sim$10$-$200 AU),
younger, giant exoplanets (and brown dwarfs). Fortunately, with the advent of the ground-based
GPI (on Gemini) (Macintosh et al. 2008), LBTI (Skrutskie et al. 2010), SPHERE (on the VLT) (Beuzit et al. 2008),
and ScExAO/Charis/HiCIAO (on Subaru) (Suzuki et al. 2010) and the space-based 
NIRCam and MIRI on JWST (Deming e tal. 2009; Shabram et al. 2009), more capable high-contrast imaging 
with near-IR performance perhaps as good as 10$^{-7}$ for $\sim$1.0 AU separations
for nearby stars ($\le$10 parsecs) will soon be routine.

\begin{figure} [h!]
 \begin{center}
    \vspace{-16pt}
     \includegraphics[width=0.40\textwidth]{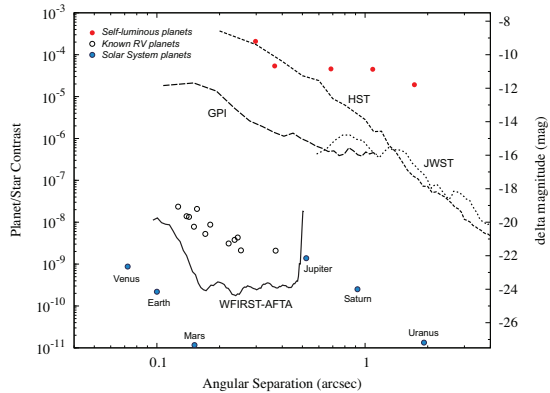}
    \vspace{-10pt}
  \end{center}
  \caption{\small Comparison of the expected planet/star contrast sensitivities of JWST, GPI, SPHERE (all in $H$ band),
   and WFIRST/CGI (in the optical).  The equivalent magnitude difference is provided on the right vertical axis.
Various comparison objects are included for reference.  Figure is taken from Spergel et al. (2013).
}
    \vspace{-10pt}
    \label{fig:1}
\end{figure}

All this recent high-contrast activity is for measurements only in the infrared.
However, albedos, phase functions, and polarization measurements in the optical 
have a long tradition in solar-system studies, and great potential to constrain 
exoplanet compositions, Keplerian elements, and aerosol and cloud properties (Marley et al. 1999).  
Hence, there is excitement surrounding the possibility that an optical coronagraph (CGI)
will be put on the Wide-Field Infrared Survey Telescope (WFIRST)/Astrophysics 
Focused Telescope Assets (AFTA) mission\footnote{One of the
stated priorities of NASA's Science Mission Directorate is to study
and characterize exoplanets. Two of the science goals highlighted in its 2014
{\it Science Plan} are to ``Explore the origin and evolution of the galaxies, stars,
and {\it planets} that make up our universe" and to ``Discover and study
planets around other stars, ..."  In the past, NASA has profitably and strategically
invested in platforms, such as HST, {\it Spitzer}, the Keck telescopes, and
{\it Kepler}, to expand our knowledge of exoplanets.  In the future, it plans
to launch JWST, and to participate in the near-- and mid-IR SPICA mission.
Moreover, the NRC's 2010 decadal survey of astronomy and
astrophysics, ``New Worlds, New Horizons" (NWNH), gave WFIRST
the highest priority for a new large space mission. In addition to the
science and payload originally envisioned for WFIRST by NWNH, NASA
is now considering an optional addition of a coronagraph
for direct imaging and low-resolution spectroscopy of giant and Neptune-like
exoplanets and brown dwarfs.  This new departure has been enabled by the
transfer from the National Reconnaissence Office
of two 2.4-meter space telescopes, one of which is
being contemplated for a WFIRST with greatly augmented
capabilities.}. It is being designed to achieve optical 
($\sim$0.4$-$1.0 $\mu$m) planet/star contrast ratios better than $\sim$$10^{-9}$ 
(Spergel et al. 2013) (with an inner-working-angle (IWA) near to or better than $\sim$0.2$^{{\prime\prime}}$) 
and some capability to measure linear polarization, though perhaps only for the imager.
Figure \ref{fig:1} compares the expected achievable planet/star contrast ratio of WFIRST/CGI
in the optical with the corresponding expected or demonstrated ratios for GPI, HST, and JWST 
in the near-infrared, and makes clear the quantum leap in capability WFIRST/CGI represents.

\bigskip

\bigskip

A low-resolution ($R = \lambda/\Delta\lambda \sim$70) IFS spectrometer is part of the baseline coronagraph design.
A jupiter-mass planet in a 4-AU orbit about a G2V star at 10 parsecs has an angular 
separation as large as 0.4$^{\prime\prime}$, and just beyond greatest elongation
(in a gibbous phase) it is expected at an age of $\sim$5 Gyrs to exhibit optical planet/star contrasts 
above $\sim$2$\times10^{-9}$.  Even at 6 AU, and at this same phase, the planet-star contrast in the
optical is expected to be roughly $10^{-9}$.  For young and/or massive giant exoplanets,
the contrasts are much higher. The optical has prominent methane features near
$\sim$0.62 \mic, $\sim$0.74 \mic, $\sim$0.81 \mic, and $\sim$0.89 \mic, ammonia has spectral bands at 
$\sim$0.65 $\mu$m and $\sim$0.79 $\mu$m, and there is a broad water band at $\sim$0.94 $\mu$m.
Rayleigh scattering off molecules and Mie-like scattering off cloud particulates and hazes
can modify planet reflectivity in diagnostic ways. A coronagraph 
on WFIRST/AFTA would be uniquely suited to provide low-resolution optical 
albedo spectra for giant exoplanets, brown dwarfs,
and some exoNeptunes at unprecedented contrast ratios.

In this monograph, we summarize the theory of the wavelength-dependent reflection albedos, phase functions, 
and polarizations as a function of 1) an exoplanet's mass, age, and composition, and 2) its Keplerian elements (including
orbital distance). This work is meant to highlight the great diagnostic potential for giant 
exoplanet characterization using the anticipated optical imaging and spectrophotometric coronagraph on WFIRST.
Low-resolution spectra and polarimetry in the optical can be used to constrain the atmospheric abundances of methane and water,
and the character, pressure levels, and physical properties of clouds and hazes expected in
giant planet atmospheres. We suggest that there should be significant variation in the signatures of different
giant exoplanets, and that Jupiter and Saturn may not be apt templates for all giant exoplanets 
at wide separations amenable to high-contrast measurement.

\section{Optical Reflection from Gas-Giant Planets: Formalism}
\label{formal}

For a planet at wide orbital separation from its parent star, the planet/star contrast ratio
in the optical and the associated geometric albedo ($A_g(\lambda)$) are the measurement goals.
They are related by the formula:

\begin{equation}
{F_p\over{F_*}} = A_g(\lambda)\left({R_p\over{a}}\right)^2\Phi(\alpha)\, ,
\label{phase_ag}
\end{equation}
where $\lambda$ is the photon wavelength, $\Phi(\alpha)$ is a function 
of orbital phase angle ($\alpha$), $R_p$ is the planet's radius, and $a$ is its orbital distance.
$A_g(\lambda)$ contains much of the information on the atmosphere's composition, pressure, and cloud/haze
profiles, though $\Phi(\alpha)$ also contains information on the atmosphere's scattering characteristics.
The scattering albedo ($\omega(\lambda)$) for single scattering off the constituents of the atmosphere is a
function of wavelength and composition and is the ratio of the scattering cross section to 
the total cross section.  For a homogeneous
atmosphere, $A_g(\lambda)$ is a direct function of $\omega(\lambda)$.  Individual scattering phase functions
for single scattering can be complicated, particularly for cloud and haze particles (for which the Mie theory
is often employed), but can generally be modeled using the Henyey-Greenstein scattering phase function:
\begin{equation}
p(\Theta) = {{1-g^2}\over {(1+g^2-2g\cos\Theta)^{3/2}}}\, ,
\end{equation}
or 
the anisotropic phase function:
\begin{equation}
p(\Theta) = 1 + 3{g}\cos{\Theta} \, ,
\end{equation}
where $\Theta$ is the single scattering angle and $g = \langle\cos{\Theta}\rangle$.
Though the phase functions for Mie scattering are more complicated,
the anisotropic phase functions above, employing only the average $g$ and 
assuming a simple cosine angular dependence,
can oftimes suffice for calculating reliable geometric albedos. 



The scalar Rayleigh phase function for scattering off individual molecules is:
\begin{equation}
p(\cos\Theta) = \frac{3}{4}(1 + \cos^2\Theta)\, .
\label{eq:rayleigh_scalar}
\end{equation}
However, vector Rayleigh scattering, with the full polarization matrix, is a more
accurate representation of Rayleigh scattering.  One difference is that, while
$A_g = 0.75$ for scalar Rayleigh scattering, it is 0.7977 for vector Rayleigh scattering,
a $\sim$6\% difference. 

A fit to $A_g$ for non-conservative vector Rayleigh scattering in semi-infinite atmospheres is:
\begin{equation}
A_g = 0.7977\frac{(1 - 0.23s)(1 - s)}{(1 + 0.72s)(0.95 + 0.08\omega)}\, , \label{eq:ag_fit2}
\end{equation}
where $s = \sqrt{1-\omega}$ (Madhusudhan \& Burrows 2011). The fit for $A_g$ is accurate to
within 1\% for $\omega \lesssim 0.99$.

Both Rayleigh and Mie scattering can result in significant polarization 
of the reflected component ($\sim$5$-$100\%), even after multiple scattering, 
suggesting that full linear Stokes polarization measurements might be useful 
for planet and planet orbit characterization (see \S\ref{pol_basics} and \S\ref{polarization}).

\begin{figure} [h!]
 \begin{center}
    \includegraphics[width=0.40\textwidth]{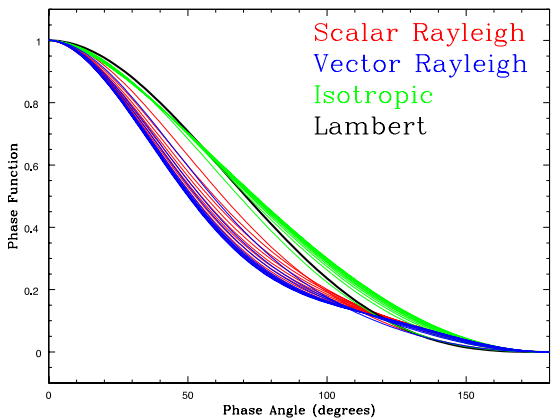}
  \end{center}
  \caption{\small Comparison of phase curves for different scattering phase functions.
The phase curves for Rayleigh scattering (both scalar and vector) and isotropic scattering
are shown for several $\omega$ values between 0 and 1; higher phase curves correspond
to larger $\omega$. For Lambert scattering, the phase curve is independent of $\omega$.
Note that the various phase curves in Figure \ref{fig:phase} can differ
between phase angles of $\sim$50$^{\circ}$ and $\sim$70$^{\circ}$ by more than a factor of two.
Figure taken from Madhusudhan \& Burrows (2011).
}
    \label{fig:phase}
\end{figure}

In any case, the net effect of an atmospheric layer on the emergent intensity is the weighted 
sum of the contributions from molecules (Rayleigh) and from aerosols of whatever kind.
The products are the measured quantities $A_g(\lambda)$ and $\Phi(\alpha)$.
Samples of derived phase functions, $\Phi(\alpha)$, for scalar and vector 
Rayleigh, Lambert reflection, and isotropic scattering are given in Figure \ref{fig:phase}.  
Inverting measurements of these quantities yields the physical and compositional characteristics 
of a planet's atmosphere, the paramount scientific goals of high-contrast imaging
in the optical.  

\bigskip

\bigskip

\subsection{Keplerian Elements}
\label{kepler}

The aspect of a planet in its orbit seen from the Earth depends upon
its Keplerian orbital elements.  In addition to the time ($t$), these are the planet's orbital angle/true anomaly 
($\theta$), orbit inclination ($i$), eccentricity ($e$), argument of periapse ($\omega_p$), 
longitude of the ascending node ($\Omega$), and period ($T$).  
\footnote{Note that $\Omega$ is necessary when a celestial (absolute) frame is designated,
(as when measuring full Stokes polarization or cataloguing many different systems in the 
same solar-system frame), but not when the system itself is allowed to provide a 
natural orientation, as in the field of radial-velocity planet
measurement.  In the latter case, $\Omega = 90^{\circ}$ can be assumed, or it can be ignored.}
The orbital inclination with respect to the reference plane of the sky
ranges from 0$^\circ$ for fully face-on orbits to 90$^\circ$ for
edge-on orbits.  The argument of periapse is the angular
distance measured along the orbit from the line of nodes to periapse, where the line of
nodes is the intersection of the sky reference plane and the orbital plane of the planet.

A planet's eccentric anomaly ($E$) can be expressed in terms of its true anomaly using:
\begin{equation}
\sin E = \frac{\sin\theta \sqrt{1-e^2}}{1+e\cos\theta}\, ,
\end{equation}
and its true anomaly is related to its mean anomaly ($M = 2\pi(t-t_p)/T$) using:
\begin{equation}
M = E - e \sin E\, .
\label{eq:kepler}
\end{equation}
Here, $t_p$ is time of pericenter passage.
We can then relate the planet's orbital angle/true anomaly ($\theta$),
as measured from periapse, to the orbit time ($t$):
\begin{equation}
t = \frac{T}{2\pi}\Bigg[2 \tan^{-1} \Bigg(\sqrt{\frac{(1-e)}{(1+e)}}\tan\frac{\theta}{2}\Bigg)-\frac{e\sin\theta\sqrt{1-e^2}}{1+e\cos\theta}\Bigg]\, .
\label{eq_time}
\end{equation}
(We have assumed that $t = 0$ at pericenter passage.)
In this way, the planetary phase curve $\Phi(\alpha)$ can be
expressed as $\Phi(t)$, obtaining a true light curve.

In general, the phase angle ($\alpha$) is related to the orbital angle ($\theta$), 
the argument of periapse ($\omega_p$), and the inclination ($i$) by,
\begin{equation}
\label{eq_elements}
\cos(\alpha) = \cos(\theta+\omega_p)\sin(i).
\end{equation}
(Here, we use a particular convention for the orbit orientation: $\theta = \alpha = 0$ when 
$\omega_p = 0$ and $i = 90^{\circ}$.) By combining eq. (\ref{eq_time}) and (\ref{eq_elements}), we obtain the exact
phase of any orbit at any time. The range of observed phase angles, $\alpha$, is given 
by $90^{\circ} - i < \alpha < 90^{\circ} + i$. Note that even when $i = 0$ the light 
curve can vary with time and phase if the orbit is eccentric.  In this case, 
$a$ (the orbital distance) changes along the orbit (eq. \ref{phase_ag}).

Though it can be presumed that the astrometry mission {\it Gaia} will have provided the 
Keplerian elements for many of the targets of WFIRST/CGI by the time it is launched,
it should be noted that both the light curves and the polarization light curves
derived using WFIRST could be used for orbit determination as well. In particular, 
the Stokes polarization vector versus time depends sensitively on the orbital 
inclination (Fluri \& Berdyugina 2010).

\subsection{Polarization Basics}
\label{pol_basics}

The Stokes vector is a four-element vector which describes the intensity and polarization
of a beam of light, and is represented as ${\textbf I} = \left[ I~Q~U~V \right]$. Here, $I$ is the total intensity, $Q$ and $U$ are
components of intensities with linear polarization, and $V$ is the intensity with circular polarization.
For exoplanets, we can generally ignore $V$.

The degree of polarization ($P$) is defined as $P = \sqrt{Q^2 + U^2}/I$. For edge-on orbits, however, the
disk-integrated Stokes $U$ vanishes due to symmetry in the north-south direction, in which case $P = Q/I$.
In general, the angle of polarization is defined as $\theta_P = \frac{1}{2}\tan^{-1}(U/Q)$.
As a generalization of eq. (\ref{phase_ag}), normalized to the stellar flux, we have for edge-on orbits:
\begin{equation}
\begin{aligned}
Q^{\prime}(\theta) & = A_g \Bigl(\frac{R_p}{a}\Bigr)^2 \Phi(\alpha) P(\theta) \, , \\
\end{aligned}
\end{equation}
where $\theta$ is the orbital phase/true anomaly, $\Phi(\alpha)$ is the phase function, and 
$P(\theta)$ is the fractional degree of polarization. $Q^{\prime}$ is defined in the orbital plane.
Since orbits are not in general edge-on, and will depend upon the orbital inclination, 
argument of perihelion, and longitude of the ascending node ($\Omega$), we need to use the mapping, 
$\cos(\alpha) = \sin(i) \cos(\theta + \omega_p)$ and rotate to the celestial frame.  
Using the M\"uller matrix for rotation through an angle $\gamma$ to transform \textbf{I} to 
the celestial frame, we obtain (Wiktorowicz 2009):

\begin{equation}
\begin{aligned}
Q(\theta) & = Q^{\prime}(\theta) \cos(2\gamma) - U^{\prime}(\theta)\sin(2\gamma)\, , \\
U(\theta) & = Q^{\prime}(\theta) \sin(2\gamma) + U^{\prime}(\theta)\cos(2\gamma)\, ,
\end{aligned}
\end{equation}
where $\gamma = \Omega + 270^{\circ}$ and we have retained the possibility that $U^{\prime}(\theta) \neq 0$.
Note again that $P(\theta) = \frac{\sqrt{(Q^2(\theta) + U^2(\theta))}}{I(\theta)}$.
Hence, with these equations we have the linear Stokes vector for arbitrary orbital elements.


\section{Albedos and Light Curves of Giant Exoplanets: General Expectations}
\label{heritage}

In the optical, from $\sim$0.4 $\mu$m to $\sim$1.0 $\mu$m, a giant exoplanet is expected to have prominent 
methane features near 0.62 \mic, 0.74 \mic, 0.81 \mic, and 0.89 \mic, a broad water feature
near $\sim$0.94 $\mu$m, and ammonia bands at $\sim$0.65 $\mu$m and $\sim$0.79 $\mu$m.
These planet/star flux ratios for distances from 1.0 AU to 10 AU
and masses from 0.5 \mj to 8 \mj can vary from above 10$^{-7}$ to $\sim$10$^{-10}$.
The strengths in absorption bands and the continuum of the reflection and albedo spectra
depend upon the atmospheric composition and molecular abundances; the temperature/pressure profile;
the possible presence of water and ammonia clouds; and the presence, altitude, complex indices of refraction,
and particle sizes of whatever aerosols and hazes may be present.  The latter could be 
polyacetylenes, tholins (Sagan \& Khare 1979; Khare et al. 1984), or various sulfur or phosphorus 
compounds.  In general, gas-phase absorption opacities increase with temperature
and metallicity, thereby decreasing the scattering albedo, and, hence, the 
geometric albedo. At the shortest wavelengths ($\le$0.5 $\mu$m), Rayleigh scattering
by gas (mostly H$_2$) and scattering by haze/cloud particles can dominate, though $A_g$ can still be 
modulated by methane features. At longer wavelengths, still shortward of $\sim$1.0 $\mu$m, molecular 
absorption bands and clouds sculpt the albedo and reflection spectra. There is often a slight 
anti-correlation in the effects of clouds in the optical and infrared, with the optical fluxes increasing
and the infrared fluxes decreasing with increasing cloud depth.

Water and ammonia clouds are expected in the atmospheres of most of the WFIRST
targets. Generally, clouds significantly increase reflection albedos.
Differences in the particle size distributions of the principal condensates
can have large quantitative, or even qualitative, effects on the resulting albedo
spectra.  In general, larger particle sizes, and wider particle size distributions
result in lower albedos.  For such condensate clouds, their physical base should be
at the ``Clausius-Clapyron" condensation level, easily determined thermodynamically.
Their geometric thicknesses are unknown, but are expected to be of order the local pressure
scale height.

In the optical, at the wide separations relevant to WFIRST/CGI, the 
planet/star flux ratio is determined almost entirely by reflected 
starlight off a giant's atmosphere.  However, such is
not the case in the infrared, where thermally re-emitted 
radiation, can determine the ratio.  This is particularly
relevant for more massive and/or young exoplanet targets for WFIRST, 
for which the outward core flux can still be large and the planet/star flux ratios 
may be thermal longward of $\sim$0.75$-$0.8 $\mu$m. Hence, although the optical spectrum
is expected to be fully phase-dependent, this may not be the case in the
near infrared, where emission may be more isotropic, depending upon the efficiency
of advection of heat to the night side of the planet. 

For instance, for Jupiter at 5.2 AU the nightside
is as bright as the dayside in the infrared longward of one micron (with an effective
temperature of $\sim$125 K).  This is in part a consequence of the shallowness of
the radiative-convective boundary and the consequent direct heating by the stellar
infrared of the convective zone itself.  Such heat is efficiently redistributed around
the planet by convection, equally to both day and night sides. This fact will 
be relevant for the WFIRST coronagraph targets. For close-in giants (i.e., transiting 
giant exoplanets not accessible to WFIRST), that interface is too deep 
to be directly heated, and as a consequence much of the stellar heat redistribution 
is done by zonal atmospheric flows that lose a large fraction of their extra heat 
before reaching the nightside. As a result, for them there are
significant day/night thermal and spectral differences. Hence, the phase curves
as a function of wavelength can be used to explore the stellar flux level 
or orbital distance at which the mechanism of day/night heat transport
for giant planets transitions.  

Though Jupiter and Saturn themselves are useful testbeds, it would be a mistake to 
extrapolate their albedo spectra to giant exoplanets as a whole.  The population
of the latter will be scattered in orbital distance, parent star, atmospheric composition,
orbital eccentricity, planet mass, and planet age.  All these characteristics, in particular
orbital distance, mass, and age, can result in large, qualitative variations in albedo
spectra among the class of giant exoplanets in the wide orbits that can be studied by WFIRST/CGI.  
For instance, the optical and geometrical thicknesses of clouds will vary with planet gravity.
Their positions in pressure space and the atmospheric temperature/pressure profiles
(that strongly influence the gas-phase molecular scattering albedos) will be functions of 
incident stellar flux, itself a function of orbital distance.  All these effects could 
translate into changes in albedo spectra of factors of a few.

The fluxes from 0.4 \mic to 0.65 \mic in the optical from irradiated, yet cloud-free, 
brown dwarfs can be enhanced by Rayleigh reflection by as much as a factor of 
10 over those from isolated brown dwarfs.  This is particularly true of old
or low-mass brown dwarfs and is a predictable function of distance (Burrows et al. 2004).

In this section, we rely heavily on the theory of giant planet albedos and light curves 
created by Sudarsky et al. (2000,2003,2005) and Burrows et al. (2004).  The models 
generated in this sequence of papers provided a useful, detailed set of theoretical models as
a function of mass, age, orbital distance, and cloud properties that can frame current
discussions and inform future models and retrievals.  

Those heritage models used an atmosphere code that placed water and/or ammonia clouds
at pressure levels determined in a fully iterative, converged fashion. They used the 
Cooper et al. (2003) theory to determine modal particle sizes, and the Deirmendjian 
(1964,1969) particle size distribution.  The modal particle size was found to be very roughly 
constant with age, but steeply decreasing with increasing surface gravity (and, hence, planet mass), and was assumed 
not to vary with altitude. The bases of the clouds were put at the intersection of the corresponding condensation 
curve with the planet's temperature/pressure profile. The scale height of a cloud 
was assumed to be equal to one pressure scale height.  For the ammonia clouds that 
form in the orbital distance sequence from 0.2 to 15 AU (i.e., at $\sgreat 4.5$ AU),
those calculations derived modal particle sizes that hovered near 50-60 \mic. The corresponding particle 
sizes in the water clouds were near 110 \mic. Little weight should be given to these 
estimates, which future data should better constrain. The irradiating stellar spectra were taken 
from Kurucz (1994) and the inner atmosphere boundary conditions were taken from the evolutionary 
models of Burrows et al. (1997). Equilibrium molecular and atomic 
compositions and abundances were obtained from Burrows \&  Sharp (1999) and the gas-phase 
opacities employed were eventually published in Sharp \& Burrows (2007).
Mie theory was used to obtain the absorptive and scattering opacities of the cloud particles.

When not plotted as a function of orbital phase, the planet/star flux ratios 
plotted below are ``phase-averaged."  By phase-averaged we mean that flux, 
when integrated over wavelength, would equal the incident 
stellar power on the planet, plus the internal luminosity of the cooling core. 
As described in Sudarsky, Burrows, \& Hubeny (2003), phase-averaging is thus a 
normalization that ensures energy conservation and is
the average flux seen at Earth over a full orbital traverse.  Hence, the maximum 
planet/star flux would always be larger than this average.


\subsection{Semi-Major-Axis/Orbital Distance Dependence}
\label{semi}

\begin{figure} [h!]
 \begin{center}
    \vspace{-16pt}
    \includegraphics[width=0.40\textwidth]{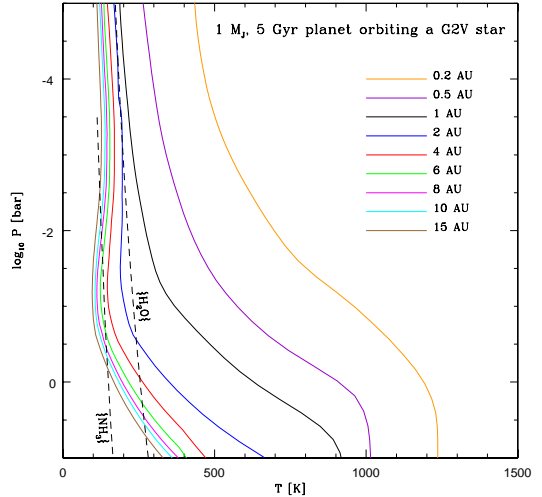}
    \vspace{-20pt}
  \end{center}
  \caption{\small Profiles of atmospheric temperature (in Kelvin) versus the logarithm base ten of the pressure (in bars)
for a family of irradiated 1-\mj giant exoplanets around a G2V star as a function of orbital distance.
Note that the pressure is decreasing along the ordinate, which thereby resembles altitude.
The orbits are assumed to be circular, the planets are assumed to
have a radius of 1.0 \rj, the effective temperature of the inner boundary flux is
set equal to 100 K,  and the orbital separations vary from 0.2 AU to 15 AU.
The intercepts with the dashed lines identified with either \{NH$_3$\} or \{H$_2$O\} denote the
positions where the corresponding clouds form. Figure taken from Burrows et al. (2004).
}
    \vspace{-10pt}
    \label{fig:TP}
\end{figure}

Figure \ref{fig:TP} depicts the resulting temperature-pressure ($T/P$) profiles for 
a 1-M$_{\rm J}$, 5-Gyr giant exoplanet orbiting a G2V star at a range of 
distances from 0.2 AU to 15 AU.  The distance determines the irradiating flux, 
which in turn, along with the interior flux, establishes the $T/P$ profile.
Also shown are the condensation curves for water and ammonia clouds.
When appropriate, and as discussed above, water and ammonia clouds are included in the models. Note that 
water clouds form around a G2V star exterior to a distance near 1.5 AU, whereas ammonia
clouds form around such a star exterior to a distance near 4.5 AU.  Consistent with these systematics, 
Jupiter itself, with a dominant ammonia cloud layer, is at the distance from the Sun of $\sim$5.2 AU.
Not shown on this figure, but a product of this study, Burrows et al. (2004) found that 
at this orbital distance and as early as $\sim$50 Myr, water clouds form in Jovian-mass 
objects.  Note also that at 4 AU, even after 5 Gyr, ammonia clouds have not yet formed 
in the atmosphere of an irradiated 1-\mj giant exoplanet.  This is not true for a similar object in
isolation (Burrows, Sudarsky, \& Lunine 2003).

\begin{figure} [h!]
 \begin{center}
    \includegraphics[width=0.40\textwidth]{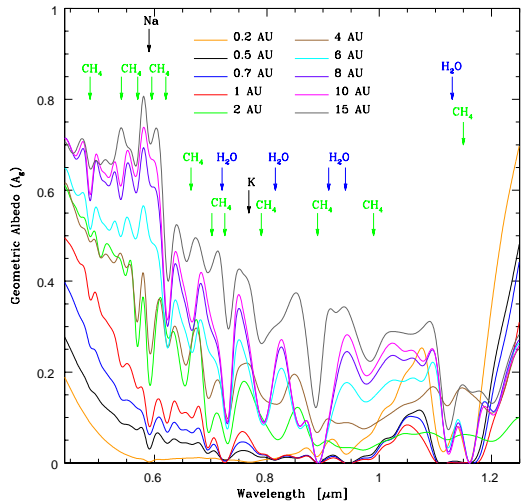}
    \vspace{-10pt}
  \end{center}
  \caption{\small Low-resolution, wavelength-dependent geometric
albedos of 1-\mj, 5 Gyr giant exoplanets ranging in orbital distance from 0.2 AU to 15 AU about
a G2V star as a function of wavelength from 0.4 to 1.25 $\mu$m.  WFIRST extends only to $\sim$1.0 \mic.
Superposed are the positions of the methane and water features in this wavelength
range.  The models have been deresolved to $R=100$. Possible reddening effects of photochemical
hazes are not incorporated. Figure taken from Burrows et al. (2004).
}
    \label{fig:albedo_seq}
\end{figure}

\begin{figure} [h]  
 \begin{center}
    \includegraphics[width=0.40\textwidth,angle = 0]{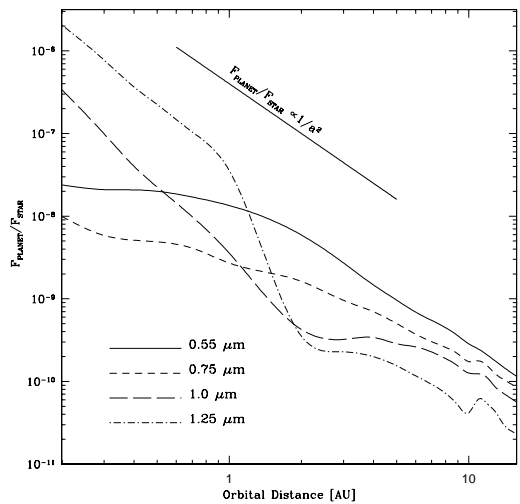}
    \vspace{-10pt}
  \end{center}
  \caption{\small Planet/star flux ratio as a function of orbital distance at 0.55 $\mu$m, 0.75 $\mu$m,
1 $\mu$m, and 1.25 $\mu$m assuming a G2V central star, taken from Sudarsky et al. (2005).  In each case, the plotted
value corresponds to a planet at greatest elongation with an orbital inclination
of 80$^\circ$.  Note that the planet/star flux ratios do not follow a simple $1/a^2$ law.
}
    \label{not_inverse_square}
\end{figure}

\begin{figure} [h!]
 \begin{center}
    \includegraphics[width=0.35\textwidth,angle = -90]{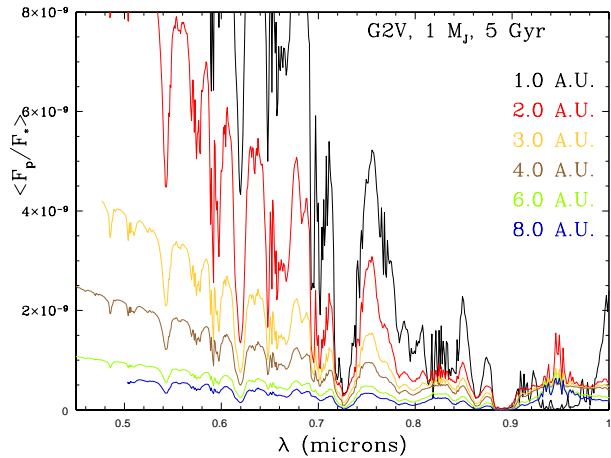}
    \vspace{-10pt}
  \end{center}
  \caption{\small Phase-averaged planet/star flux ratios in the optical (0.45 to 1.0 $\mu$m)
for a 1-M$_{\rm J}$ giant planet orbiting a G2 V star as a function of orbital distance (semi-major axis)
from 1.0 to 8.0 AU  The model data were taken from Burrows et al. (2004).  Note that some models
seem to show emission, some absorption, at the water feature near 0.94 $\mu$m. See text for a discussion.  
These models employ the geometric albedos shown in Figure \ref{fig:albedo_seq}, but include a thermal component
due to the residual cooling of the planet core (Burrows et al. 1997).
}
    \label{distance}
\end{figure}

Since the atmospheric temperatures vary with stellar distance, because such variation determines
whether water or ammonia clouds form, the albedos will vary significantly with orbital distance.
In fact, one expects the albedo spectrum of a giant planet to depend sensitively upon orbital 
distance, mass, age, and parent star. Figure \ref{fig:albedo_seq} depicts the geometric albedo
spectra of a 1$-$M$_{\rm J}$, 5-Gyr exoplanet at the various orbital distances used in Figure \ref{fig:TP}.  
Also shown are the positions of the prominent molecular bands of methane and water.
The presence of clouds profoundly affects the reflectivity of the planet. Figure \ref{fig:albedo_seq} demonstrates
that the closest, being hotter and clearer, have low albedos, while the furthest, being cooler
and having either water clouds ($\le$300 K; generally $a \ge 1.5$ AU) or ammonia
clouds ($\le$160 K; generally $a \ge 4.0$ AU), have higher albedos. 
The optical albedo rises significantly with increasing orbital radius between
$\sim$ 0.2 AU and 1 AU, even though there are no water clouds
present.  With the onset of water clouds, the optical albedo rises further
still, which is seen clearly in the model at 2 AU.  The onset and thickening
of reflective ammonia clouds at larger orbital distances results in the
highest albedos at most optical and near-infrared wavelengths.  
For greater distances than $\sim$1.0 AU, in these models the presence of water 
clouds slightly mutes the variation with wavelength in the planetary spectra.
Hence, smoothed water features and methane bands predominate beyond $\sim$1.5 AU.
Note that because the methane abundances and cloud presence and properties are functions of temperature,
the albedo spectrum is a non-monotonic function of planet mass and age (which determine in
part the degree to which the planet has cooled), orbital distance, stellar type, and planet
elemental composition.  Moreover, as Figure \ref{not_inverse_square} demonstrates, 
the planet/star flux ratio does not follow an inverse-square law with orbital distance,
and the deviations from such behavior are functions of wavelength.  
Hence, giant planet reflectivity is a rich function of planet/orbital properties and its 
optical spectrum can in principle elucidate the planet's atmosphere and history.
Figure \ref{distance} provides some of the phase-averaged planet/star flux ratios
from $\sim$0.4 \mic to $\sim$1.0 \mic that correspond to the models in Figures 
\ref{fig:TP} and \ref{fig:albedo_seq}.

\begin{figure} [h!]
 \begin{center}
    \includegraphics[width=0.35\textwidth]{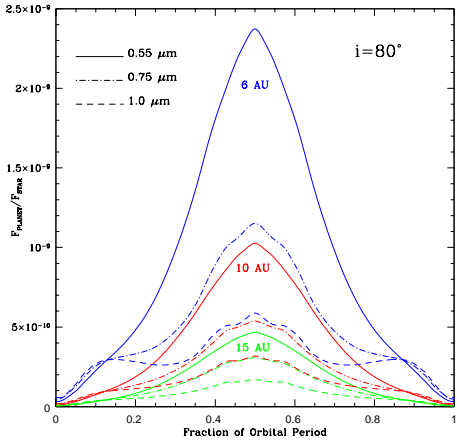}
  \end{center}
  \caption{\small Light curves at 0.55 $\mu$m,
0.75 $\mu$m, and 1 $\mu$m for 1-M$_{\rm J}$ giant exoplanets
in circular orbits inclined to 80$^\circ$ at distances of 6 AU, 10 AU,
and 15 AU from a G2V star.  The planet/star flux ratio
is plotted.  Each of these models contains an ammonia ice cloud layer above a deeper water
cloud deck. Taken from Burrows et al. (2004) and Sudarsky et al. (2005).
}
    \label{fig:outer_inner1}
\end{figure}

\begin{figure} [h!]
 \begin{center}
    \includegraphics[width=0.35\textwidth]{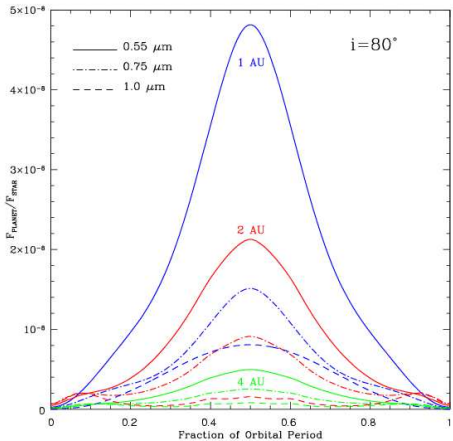}
  \end{center}
  \caption{\small Light curves at 0.55, 0.75, and 1.0 microns for model gas giants
in circular orbits inclined to 80 degrees at distances of 1 AU, 2 AU, and 4 AU
from a G2V star. The planet/star flux ratio is plotted.  The
models at 2 AU and 4 AU contain water ice clouds in their upper atmospheres,
while the 1 AU model does not. Taken from Burrows et al. (2004) and Sudarsky et al. (2005).
}
    \label{fig:outer_inner2}
\end{figure}

Figures \ref{fig:outer_inner1} and \ref{fig:outer_inner2} incorporate phase functions ($\Phi(\alpha)$)
calculated by Sudarsky et al. (2005) and Burrows et al. (2004) and convey representative theoretical 
phase dependences/light curves for semi-major axes of 6, 10, and 15 AU and of 1, 2, and 4 AU, respectively, at 0.55 \mic, 
0.75 \mic, and 1.0 \mic.  An orbital inclination of 80$^{\circ}$ is assumed. These figures also provide the
maximum planet/star ratios at greatest elongation (its maximum angular separation
as seen from Earth).  Given a stellar distance and an inner working 
angle (IWA $\sim$0.1$-$0.2$^{\prime\prime}$), Figures \ref{fig:outer_inner1} and 
\ref{fig:outer_inner2} can be used to compare with the expected contrast ratio performance of WFIRST/CGI
to determine detectability as a function of phase and orbit.  As these figures indicate, WFIRST 
should have sensitivity to a wide range of 1-\mj giant exoplanets orbiting a solar-like star
out to tens of parsecs. 

As stated in Sudarsky et al. (2005), at 0.55 $\mu$m and 0.75 $\mu$m, the planet/star flux ratio near 
full phase is 3 to 4 times its value at greatest elongation.  Hence, at 
these wavelengths, the planet/star flux ratios are not very
different from those that would arise if we were to assume a
Lambert phase function, in which case the ratio at full phase
would be $\pi$ times larger than at greatest elongation.
However, at other phase angles and other wavelengths, model
phase functions frequently differ by a wider margin from
the Lambert phase function.

\begin{figure} [h!]
 \begin{center}
    \vspace{-10pt}
    \includegraphics[width=0.35\textwidth]{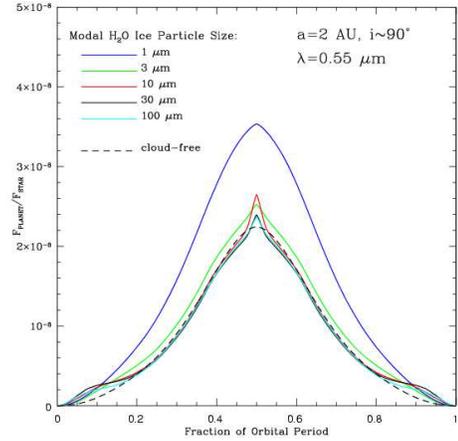}
    \vspace{-10pt}
  \end{center}
  \caption{\small The dependence of the planet/star flux ratio on condensate particle
size at a wavelength of 0.55 microns.  Model light curves for EGPs
at 2 AU with modal H$_2$O ice particle sizes of 1, 3, 10, 30, and 100 microns
are depicted. Shown for comparison is a cloud-free model (black dashed curve).
In order to show the full variation in the shapes and magnitudes of the light
curves with particle size, we have set the orbital inclination to
approximately 90 degrees so that the opposition effect, present for many
of the models, can be seen in full.  Transit effects are not modeled.
Figure taken from Sudarsky et al. (2005).
}
    \vspace{-10pt}
    \label{fig:inner_size1}
\end{figure}

\begin{figure} [h]  
 \begin{center}
    \includegraphics[width=0.35\textwidth]{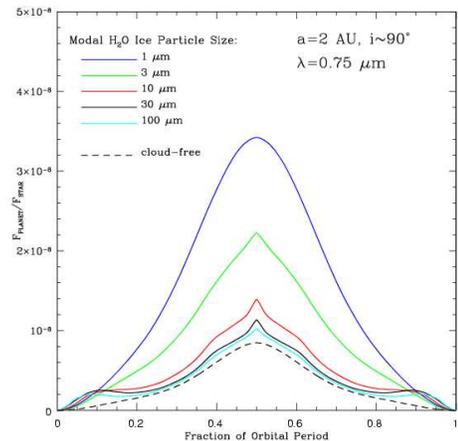}
    \vspace{-10pt}
  \end{center}
  \caption{\small Same as in Figure \ref{fig:inner_size1}, but for 0.75 $\mu$m.
}
    \label{fig:inner_size2}
\end{figure}

To gauge the dependence of the planet/star flux ratios on the cloud
particle size assumed, Figures \ref{fig:inner_size1} and \ref{fig:inner_size2} (taken from Sudarsky et al. 2005)
explore this dependence for water clouds in a 1-\mj model at 2 AU from a G2V star at two 
different wavelengths.  Also shown is the corresponding cloud-free model\footnote{Recall that the 
models presented via Figures \ref{fig:TP}, \ref{fig:albedo_seq}, and \ref{distance}
employed the model of Cooper et al. (2003) to obtain the modal particle size, and that for a water
cloud such particle sizes varied around $\sim$100 \mic.}.
At both 0.55 $\mu$m and 0.75 $\mu$m, the 1 $\mu$m particle size models
exhibit higher planet/star flux ratios and smoother light curves than
for those models with larger particles.  At 0.55 $\mu$m, the planet/star flux ratios do not
vary substantially over a range of modal particle sizes from 3 to 100 $\mu$m
(although the shapes of the light curves differ somewhat).  This result contrasts with
those of the same giant at 0.75 $\mu$m, where the planet/star flux ratio
becomes progressively lower with increasing particle size.  Such
results indicate the importance not only of condensate particle size, but
of wavelength-dependent light curves.

\subsection{Mass and Age Dependence}
\label{mass}

\begin{figure} [h!]
 \begin{center}
    \includegraphics[width=0.35\textwidth,angle = -90]{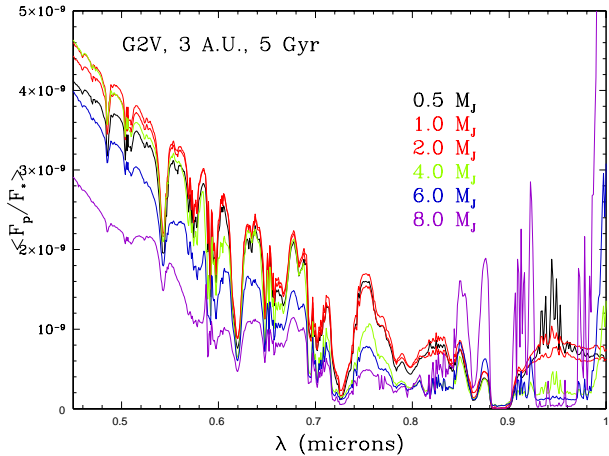}
    \vspace{-10pt}
  \end{center}
  \caption{\small Phase-averaged planet/star flux ratios in the optical
as a function of giant planet mass (in M$_{\rm J}$) at 3 AU from a 5 Gyr-year-old
G2 V star.  Higher gravity generally leads to lower fluxes at short wavelengths.
Note that for these mass/age combinations, the more massive planets (6.0 and
8.0 M$_{\rm J}$) have enhanced fluxes longward of $\sim$0.8 $\mu$m, due to
a competitive contribution of thermal emission powered by their greater
residual heat of formation (Burrows et al. 1997). 
This variety highlights the diagnostic potential of even 
low-resolution optical spectra and spectrophotometry to
constrain various giant planet properties.
}
    \label{mass_plot}
\end{figure}

Figure \ref{mass_plot} shows the predicted planet/star phase-averaged contrast ratio
spectrum for giant planet masses from 0.5 to 8.0 \mj at an age of 5 Gyr and at an orbital distance of 3 AU. 
Older giant exoplanets have progressively deeper water clouds.  These older clouds are also
thicker (have higher column depths).  This results in a slightly
increasing reflected optical flux with increasing age.  Conversely, larger-mass 
giants have higher surface gravities, which result in water clouds
with lower column depths, and, hence, lower optical depths, despite the possible
contrary trend of modal particle size. Therefore, as seen in Figure \ref{mass_plot}, at a given 
age and at the shorter optical wavelengths, higher-mass giant 
exoplanets have slightly smaller planet/star flux ratios. However, also 
at a given age, higher mass giant planets have higher \teff\ and internal luminosities.
This is due to the higher core fluxes of more massive planets that are still cooling
from formation.  Therefore, at longer wavelengths (greater than or equal to $\sim$0.8 \mic) still accessible to
WFIRST/CGI, the flux ratios at a given age are higher for more massive giants. 
For such planets younger than 5 Gyr, the effect can be larger still.
Therefore, for more massive giant planets and brown dwarfs, and for their younger cohort,
the flux ratios longward of 0.7$-$0.8 \mic can be much higher than for an old, 1-\mj giant such as Jupiter.
This also means that the planet/star flux ratios at these longer wavelengths for the more
massive and/or younger exoplanets will be weaker functions of orbital phase and their apparent albedos
can be greater than one. For such giant exoplanets, extrapolations from Jupiter 
should be wildly in error.

Around 0.94 $\mu$m, as seen in Figure \ref{mass_plot}, some of the models show flux enhancements,
while others show the flux deficits one would expect due to band absorption.  This is 
due in part to a combination of the dependence of the water absorption opacity
on temperature and on metallicity.  All else being equal, the higher the metallicity
or the higher the temperature the lower the water band peak in the albedo spectrum. 
Condensation of gaseous water at lower atmospheric temperatures will also affect 
the gas-phase water absorption opacity.  All these effects alter the scattering 
albedo in this wavelength band, and, hence, the associated geometric albedo.  
The magnitude of this effect will depend upon details, but its presence 
could constrain 1) metallicity and 2) water temperature and 
condensation.  Hence, the 0.94-\mic water band is expected to be a diagnostic 
of metallicity (or elemental composition), orbital distance, and mass. 

\subsection{Predictions for Known Radial-Velocity Giant Exoplanets}
\label{known}

\begin{figure} [h!]
 \begin{center}
    \includegraphics[width=0.35\textwidth,angle = -90]{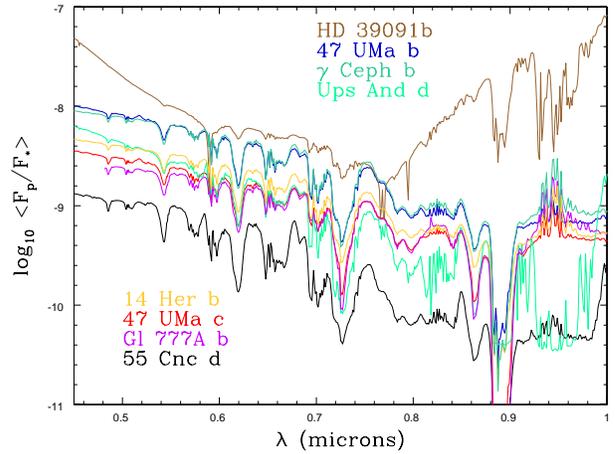}
    \vspace{-10pt}
  \end{center}
  \caption{\small Phase-averaged planet/star flux ratios in the optical
for representative non-transiting radial-velocity exoplanets with planet-star
angular separations accessible to WFIRST/CGI at greatest elongation.
These include 55 Cnc d (0.44$^{\prime\prime}$), 47 UMa c (0.28$^{\prime\prime}$),
47 UMa b (0.16$^{\prime\prime}$), Ups And d (0.19$^{\prime\prime}$),
Gl 777A b (0.23$^{\prime\prime}$), HD 39091b (0.16$^{\prime\prime}$),
14 Her b (0.15$^{\prime\prime}$), and $\gamma$ Ceph b (0.15$^{\prime\prime}$).
The angles in parentheses are the ratios of semi-major axis to Earth-star distance.
These models are taken from Burrows et al. (2004). 
Notice that some of the models have local peaks (e.g., for $\gamma$ Ceph b), some local 
troughs (e.g., for Ups And b), in a 0.94 $\mu$m water feature.  This can be indicative
of the presence or absence of gaseous water in abundance and of the atmospheric 
temperature (which affects the corresponding absorption opacity and scattering albedo, $\omega$).  
Also, as the model for HD 39091b indicates, massive giant planets will retain their
heat of formation and emit copiously in the 0.7 $-$ 1.0 $\mu$m band, far more than
they reflect. Hence, this spectral region is a good mass/age index of the orbiting giant.
As this plot demonstrates, F$_p$/F$_*$ is not merely a scaled one-parameter family
of passive and uniform reflection spectra, but has many diagnostic band ratios
and/or colors in the optical.
}
    \label{specifics}
\end{figure}

One set of useful and reliable targets for a WFIRST exoplanet campaign will
be nearby known RV planets at wide separations. There are perhaps $\sim$20$-$30
of these, and they include 55 Cnc d (IWA(max) = 0.44$^{\prime\prime}$),
47 UMa c ((IWA(max) = 0.28$^{\prime\prime}$), 47 UMa b (IWA(max) = 0.16$^{\prime\prime}$),
Gl 777Ab (IWA(max) = 0.23$^{\prime\prime}$), HD 39091b (IWA(max) = 0.16$^{\prime\prime}$),
14 Her b (IWA(max) = 0.15$^{\prime\prime}$), and $\upsilon$ And d
(IWA(max) = 0.19$^{\prime\prime}$). Table \ref{data.specific} lists some of these planets and
some of their relevant characteristics. Phase-averaged planet/star ratios
versus wavelength for this target class are shown in the left panel of Figure \ref{specifics}. As these figures
demonstrate, the flux ratios in the optical for these planets are in principle accessible to
WFIRST and vary substantially from object to object. We note that the high mass of HD 39091b
elevates its likely flux ratio longward of $\sim$0.75 \mic by an order of magnitude or more,
bearing out the point made about younger and more massive exoplanets at longer wavelengths
made in \S\ref{mass}.

\subsection{Inclination and Eccentricity Dependence}
\label{eccen}

\begin{figure}
 \begin{center}
    \includegraphics[width=0.4\textwidth]{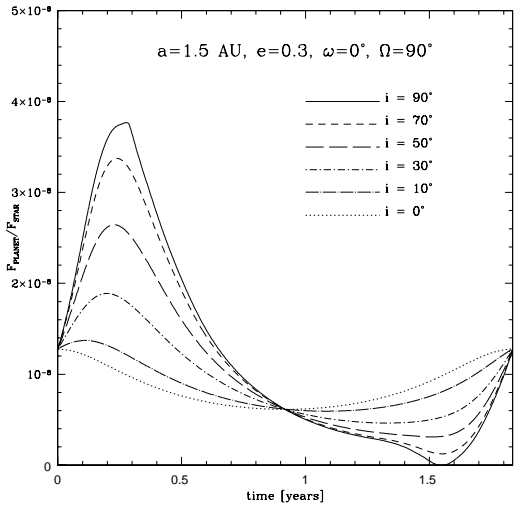}
    \vspace{-10pt}
  \end{center}
  \caption{\small Variation with inclination of the optical light curve for an elliptical orbit
(G2V central star, $a$ = 1.5 AU, $e$ = 0.3).  For a highly-inclined orbit, the
peak of the planet/star flux ratio is a factor of $\sim$3 greater than for a
face-on ($i=0$ degrees) orbit.  The (symmetric) variation for the face-on case is due
entirely to the variation in the planet-star distance over an eccentric orbit.
Taken from Sudarsky et al. (2005).
}
    \label{fig:inclination}
\end{figure}

\begin{figure} [h!]
 \begin{center}
    \includegraphics[width=0.40\textwidth]{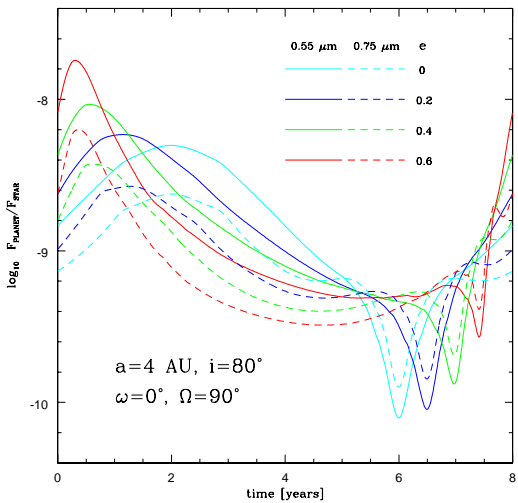}
    \vspace{-10pt}
  \end{center}
  \caption{\small The optical planet/star flux ratio as a function of eccentricity
for a cloud-free 1-M$_{\rm J}$ giant planet at 1 AU, fixing $i$ at $90^\circ$
and $\omega_p$ at $0^\circ$.  Only for $e$=0 does the
peak of the light curve coincide with full phase. Taken from Sudarsky et al. (2005).
}
    \label{fig:eccentricity}
\end{figure}

The interpretation of optical reflection spectra is complicated
by the need to factor in the effects of orbital inclination, eccentricity,
and argument of periapse $-$ planet orbits will not in general be circular and
edge-on. In addition, the planets will not always be at greatest elongation.
In principle, the Keplerian elements can be determined independently
from RV data, when available, but can also be derived astrometrically
from multi-epoch WFIRST observations themselves. Since {\it Gaia} astrometry
on the observed systems will be public by the time WFIRST is operational,
that source of orbital information could be invaluable.  Be that as it may,
as stated, most planets observed by the WFIRST coronagraph will not be in simple, edge-on,
circular orbits and the corrections, though straightforward theoretically,
are not small.  We recall that Figure \ref{fig:phase} provided some model phase 
functions and noted that at a given phase or mean anomaly $\Phi(\alpha)$  
can range by factors of two or more.  Moreover, the Lambert phase function 
frequently used can be significantly in error. Despite this, the phase variation 
of the intensity should be a smooth function of orbital phase, while we expect that 
the variation of the planet's polarization and polarization angle could be much less smooth.
Fluri \& Berdyugnia (2010) suggest, and experience with Venus demonstrates 
(Hansen \& Hovenier 1974), that polarization as a function of phase (and wavelength) 
can be used to constrain not only cloud properties, but orbital elements as well (\S\ref{polarization}).

Figure \ref{fig:inclination} depicts an example of
planet/star flux ratio variation with orbital inclination (for an eccentric orbit),
and Figure \ref{fig:eccentricity} depicts sample variations with eccentricity.  Clearly, large variations in
flux ratio along an orbital traverse are possible and need to be incorporated into the
modeling, particularly of the absolute flux levels.  What these figures show is 1) that
the flux ratios anticipated are indeed within WFIRST's coronagraphic reach, and 2)
that the dependence upon Keplerian elements must be incorporated into the analysis.

\section{Jupiter's Albedo Spectrum}
\label{jupiter_albedo}

\begin{figure}
 \begin{center}
    \vspace{-10pt}
    \includegraphics[width=0.30\textwidth,angle=-90]{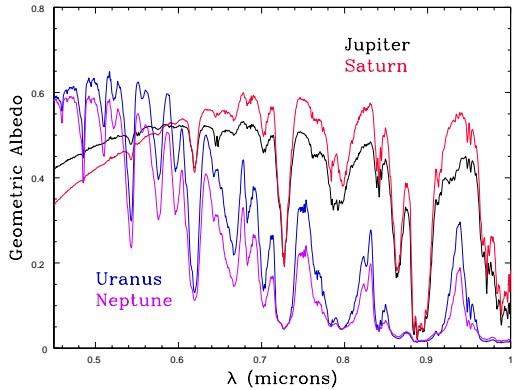}
  \end{center}
  \caption{\small The geometric albedo spectra of Jupiter, Saturn, Uranus and Neptune,
from Karkoschka 1994.  Note the dominance of the methane bands, but also the differences between
Jupiter and Saturn and Uranus and Neptune.  The greater depths of the features for the ``ice giants"
are due to the greater column depths for the lower gravities and the higher methane abundances. 
}
    \label{solar_giants}
\end{figure}

Even if present in trace amounts, stratospheric hazes 
and tropospheric ``chromophores" (West et al. 1986; Pope et al. 
1992) can alter emergent reflection spectra 
substantially, particularly in the ultraviolet and blue regions of the 
spectrum. Karkoschka (1994,1998) has obtained full-disk geometric albedo 
spectra of Jupiter, Saturn, Uranus, and Neptune.  Figure \ref{solar_giants} portrays
these optical albedo spectra, which show that the atmospheres of Jupiter and Saturn, 
but not of Uranus and Neptune, indeed contain a trace species well-mixed in their upper 
ammonia cloud decks that is depressing the reflected flux shortward of 
$\sim$0.6 \mic from what one would expect from the increase with frequency of 
the Rayleigh and ammonia cloud scattering cross sections (\S\ref{heritage}). 
It is this trace species, with a mixing ratio of perhaps no more
than 10$^{-9}$$-$10$^{-8}$, that gives Jupiter and Saturn their reddish hues,
and it is its absence in Uranus and Neptune that maintains their expected bluish tinge.
The optical albedo spectra longward of $\sim$0.6 \mic of these four solar-system 
planets are otherwise as expected, given the systematics described in \S\ref{heritage}, 
and are dominated by methane absorption features and ammonia clouds. 

\begin{figure}
 \begin{center}
    \vspace{-10pt}
    \includegraphics[width=0.40\textwidth,angle=0]{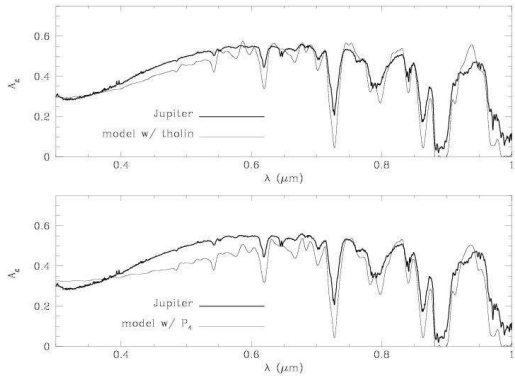}
  \end{center}
  \caption{\small The full-disk geometric albedo spectrum 
of Jupiter (thick curve) from Karkoschka 1994, compared with model albedo spectra (thin
curves).  The top model utilizes tholin as a chromophore,
while the bottom model uses P$_4$.  Models taken from Sudarsky et al. (2000).
}
    \label{jupiter_plot}
\end{figure}

Figure \ref{jupiter_plot} depicts two model comparisons with Jupiter's
measured albedo spectrum.  Relying upon Mie scattering theory and choices for chromophore
particle size distributions, tholin appears to reproduce the UV/blue
region of the albedo better than P$_4$, but a number of other possible trace 
species are still in contention. Polyacetylenes (Rages 1997), tholins, sulfur compounds, 
or elemental phorphorus have been suggested, and UV photolytic chemistry may be involved.

Given this mixed picture, it is not known how prevalent this trace
species may be among the giant exoplanets.  As in the case of 
Uranus and Neptune, they may not be present generically. Be that as it may, libraries
now exist for the complex indices of refraction of many potential chromophores,
and these can be used, if necessary, to fit the anticipated data shortward of $\sim$0.6 \mic.
Note that the mere change of average albedo slope has information $-$
correlations of the red or blue colors of WFIRST/CGI targets with stellar type and metallicity, planet mass,
or orbital distance could teach us not only about giant exoplanets, but about the solar-system giants
as well. 

In any case, the properties of clouds and hazes will have to be parametrized when fitting
the WFIRST/CGI albedo data. Physically, the cloud/haze position, abundance, modal particle size, and complex
index of refraction as a function of wavelength are determinative.  Practically, the position
of the cloud top, the particle scattering albedo,
particle number density, and wavelength dependence will have to be modeled. Since clouds
and hazes introduce a continuum extinction spectrum, this can be modeled with only a
few degrees of freedom.  The indices of water and ammonia droplets are understood, and the positions
of their cloud bases can be derived from their known condensation curves, but specific
haze materials can be injected into the fits, or parametrized.  We note that the various methane
spectral features are formed at different heights and pressures in the atmosphere, providing a
means to break possible (quasi-)degeneracies and derive cloud and approximate molecular abundances 
separately. Hence, with the presence of clouds, but also of multiple bands, fitting multiple 
methane and water feature depths simultaneously should allow one to break the various degeneracies 
and derive physical abundances, gravities, and, perhaps, temperatures.

\section{Polarization}
\label{polarization}

\begin{figure}
 \begin{center}
    \includegraphics[width=0.40\textwidth]{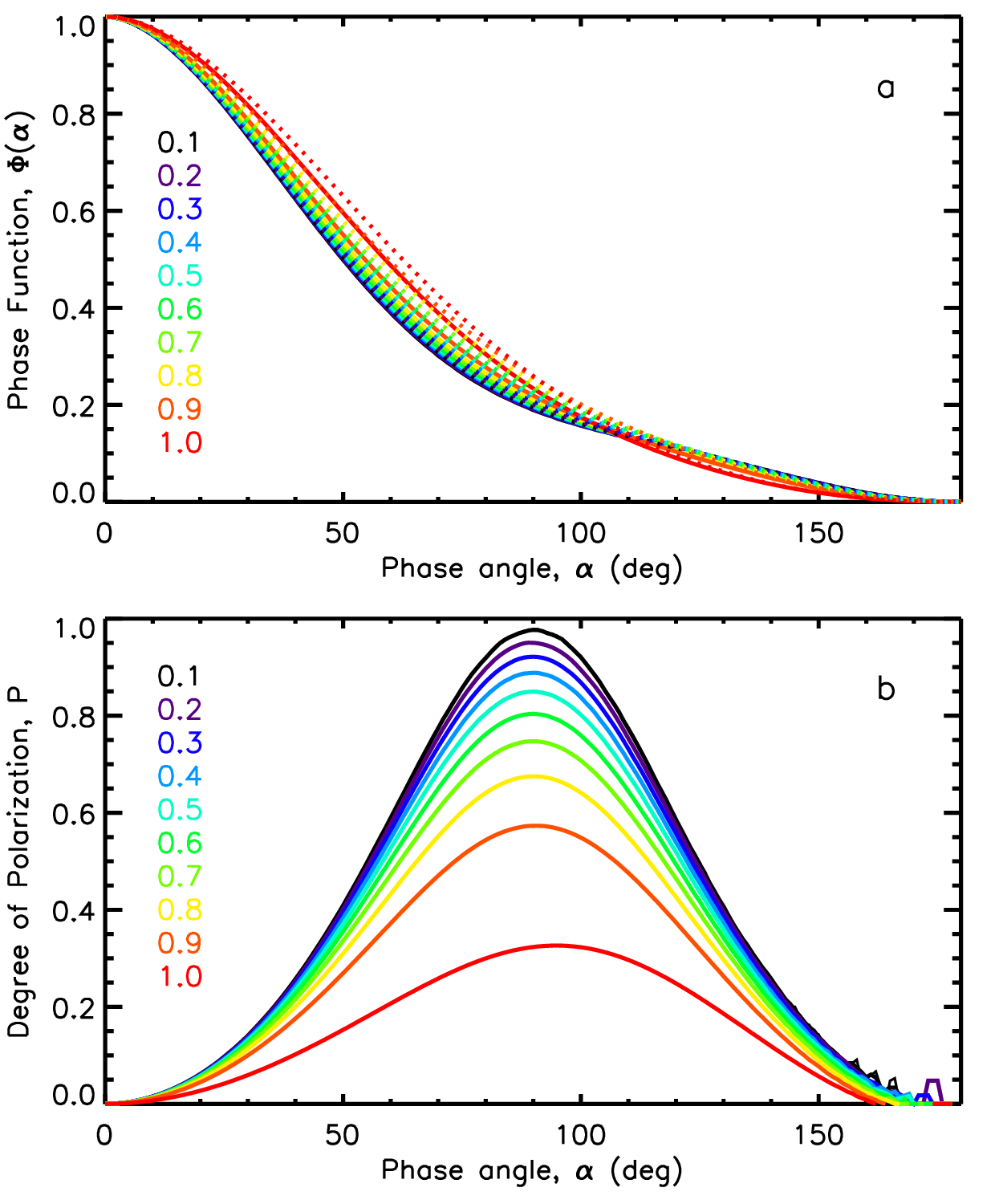}
    \vspace{-10pt}
  \end{center}
  \caption{\small The various phase functions for a vector Rayleigh-scattering atmosphere (top), and the corresponding
linear polarizations (bottom) as a function of orbital phase and scattering albedo ($\omega$, column of numbers) 
(from Madhusudhan \& Burrows 2011).
}
    \label{madhu_pol}
\end{figure}

  The exoplanet case for having a full linear Stokes capability on
WFIRST/CGI is strong (Schmid et al. 2006).  
Estimates of peak exoplanet polarizations, with
either Rayleigh or various Mie particles, range from 10\% to 90\% 
(Madhusudhan \& Burrows 2011; Stam et al. 2004). 
The bottom panel of Figure \ref{madhu_pol} plots the magnitude of the polarization and
its variation with orbital phase, as a function of particle scattering
albedo ($\omega$), due to vector Rayleigh scattering in a homogeneous
atmosphere.  Table \ref{vec_ray} provides the geometric albedos, peak polarizations, and
phase angle at peak for vector Rayleigh scattering, also as a function of $\omega$.
For such an atmosphere, the peak polarization is always above 32.6\%. Haze and cloud 
particles (as determined from Mie theory) are generally less polarizing than molecular 
Rayleigh scatterers, but are Rayleigh-like in the small particle ($x = 2\pi a/\lambda << 1$) limit.
This difference between the polarization for gas molecules and that for typical
cloud and aerosol particles makes linear polarization a valuable
tool for determining the presence and nature of those particles (Hansen \& Travis 1974). This is
all the more germane when we note that at shorter wavelengths,
polarization generally contains information about the upper atmosphere,
while at longer wavelengths clouds deeper in the atmosphere are probed.
For example, if there is a Rayleigh scattering layer at depth, the more penetrating long-wavelength
radiation can emerge more polarized than the short-wavelength radiation.  Hence, a 
polarization color index (for example, ratioing blue with red or red with I-band) can reveal 
the presence, thickness, and optical depth of a Rayleigh scattering layer (Buenzli \& Schmid 2009).
Moreover, the total intensity generally decreases with increasing orbital phase angle from 0 to 90$^{\circ}$,
while the percent polarization generally increases in this interval.  This results in a phase angle
at which the polarized flux (as opposed to the polarized fraction) is a maximum.  For conservative 
Rayleigh scattering, this angle is $\sim$65$^{\circ}$, but it is generally larger for haze particles.
Hence, this angle alone can be used as a diagnostic of the presence of aerosols (Buenzli \& Schmid 2009).

\begin{figure} [h!]
 \begin{center}
    \includegraphics[width=0.45\textwidth]{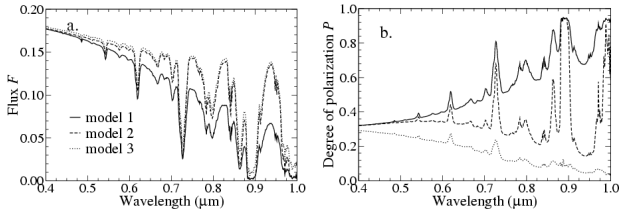}
    \vspace{-10pt}
  \end{center}
  \caption{\small Multi-wavelength optical polarimetry measurements can constrain
the atmospheric properties of giant exoplanets. The flux (left, in arbitrary units) and degree
of polarization (right) of starlight reflected by three Jupiter-like gas
giants for $\alpha = 90^\circ $. Planetary model atmosphere 1 (solid lines)
contains only molecules, model 2 (dashed lines) is similar to model 1, except
for a tropospheric cloud layer, and model 3 (dotted lines) is similar to model 2,
except for a stratospheric haze layer. Figure taken from Stam et al. (2004).
}
    \label{stam}
\end{figure}

Importantly, multiple scattering does not smooth out
angular features in the polarization (i.e., $Q$), as it does in the light
curve intensities (i.e., $I$).  Those features bespeak particle refractive
indices and sizes. Indeed, Stam et al (2004) note that planetary polarization at a given wavelength 
may be a more diagnostic function of phase than the light curve (i.e., $\Phi(\alpha)$) in probing the 
cloud and haze structures and positions in gas giant atmospheres.  They emphasize the relative 
nature of a polarization measure (hence, its intrinsically more accurate 
character) and conclude that the degree of polarization in both spectral bands 
and the continuum depends strongly upon atmospheric composition and structure.
More powerfully, when combining geometric albedo and polarization data as a function of wavelength
simultaneously (for a known orbital phase), one is able to break various approximate degeneracies
introduced by the presence of clouds/hazes.  Figure \ref{stam} from Stam et al. (2009)
depicts both a scaled flux (left) and the degree of polarization (right) as a function of
wavelength in the optical for three different atmosphere models at quadrature.  For all models, the methane
spectral features are prominent, and for the cloudy models the fluxes are larger than for
the pure Rayleigh scattering model.  At the same time, the polarizations for the cloudy models are smaller,
and have the opposite dependence upon wavelength, particularly in the continuum.  Moreover,
the tropospheric and stratospheric cloud models can be distinguished by the line to continuum
polarization ratios.  Finally, though the fractional polarization can be large in strong absorption bands,
while the flux itself is small (due to the absorption), the continuum does not suffer from this effect.
Hence, both flux and polarization ratio indices between the absorption bands and the continuum
are useful diagnostics of scattering cloud depth, the presence of aerosols, particle size and make-up, 
and the relative contribution of Rayleigh scattering.

Fluri \& Berdyugina (2010) have suggested, if a reasonable partial phase curve can 
be obtained, that Keplerian orbital elements of an exoplanet such as inclination, 
eccentricity, and the position angle of the ascending node can be constrained
using polarization measurements as a function of orbital phase or mean/true anomaly. 
The peak polarization at a given wavelength is independent of orbital inclination 
and reflects the presence and layering of hazes and clouds, but the polarization 
phase curve can constrain orbital elements such as inclination particularly well (\S\ref{pol_basics}).  
We note, however, that {\it Gaia} astrometry may have already provided the orbital elements of many of
the WFIRST exoplanet targets prior to the latter's launch, in which case the orbital information extracted from
the rotation of the direction of polarization with phase would be, however useful, merely confirmatory. 
In fact, the likely availability of planet orbital data from {\it Gaia} will make interpreting 
both intensity and polarization data much less ambiguous for single-epoch observations by
providing the planet's Keplerian elements.  Knowledge of the orbit would usefully reduce
the number of free parameters to be deduced.

As Hansen \& Hovenier (1971) have articulated, the sharpness of the angular variations 
in the polarization increases markedly for larger particles or shorter wavelengths 
(larger $x = 2\pi a/\lambda$), and this can be used to directly constrain particle size.
In addition, if there are multiple particle species, each particle type contributes
to the polarization approximately according to its share of the ``single" scattered light (Hansen \& Hovenier 1974).
The robustness of the qualitative single-scattering result in the context even of multiple scattering
leaves open the possibility that spherical droplets could imprint rainbow and/or glory effects.
Interestingly, rainbow angles were seen in the polarization measurements of Venus. These 
measurements from 0.365 \mic to 0.99 \mic as a function of
Venus' orbital phase revealed a mean droplet size of 1.05 $\mu$m (with a
$\sim$7\% logarithmic spread), with a real part of the index
of refraction from 1.46 to 1.43 for wavelengths from 0.365 to  0.99$\mu$m (Hansen
\& Hovenier 1974). The latter cinched the composition of its clouds to be sulfuric acid.
The polarization for Venus at 0.55 \mic is $\sim$5\% and the Rayleigh fraction was determined 
to be $\sim$4.5\%.  Multi-band optical polarization measurements constrained Venus'
cloud modal particle size to better than 10\%, and the real part of the
index of refraction to $\sim$1\%.  

In addition, as measured by Pioneer 11 (Tomasko \& 
Smith 1982) and Voyager 2 (West et al. 1983) at quadrature, the fractional polarization of 
the hazy moon Titan is near $\sim$50\% from the R band to the near UV. 
Such a large polarization confirmed the small size of the aerosol particles 
in Titan's upper atmosphere, but the concomitant large forward scattering inferred
as a function of phase angle (Rages, Pollack, \& Smith 1983) suggested that the 
haze particles could not be spherical (West \& Smith 1991). Hence, light curve and polarization 
data together revealed quite a bit about the physical character of the particles in Titan's haze layer.
However, despite the inherent usefulness of polarization measurements as a function
of wavelength and phase, few polarization measurements have been made of
the outer-solar-system giants.  Exceptions are the Pioneer 10 and 11
photopolarimetry measurements in the red and blue of Jupiter and Saturn 
(Smith \& Tomasko 1984).  These measurments were, however, angularly 
resolved and not full-disk.  One reason for such sketchy full-disk coverage is that the 
two polarization channels on Voyager 1 failed after launch, depriving science
of such useful data. Another is that the phase angles of Jupiter and
Saturn at Earth are a paltry $\sim$11$^{\circ}$ and $\sim$6$^{\circ}$,
respectively. A third is the priority placed by planetary scientists on
in-situ composition, charged-particle, and magnetic-field probes. However,
since such in-situ measurements are not possible for exoplanets and the
entire focus of exoplanet research is on remote sensing, a polarization
capability should seem compelling.  Hence, there is a strong 
case for full-linear Stokes imaging polarimetry on WFIRST-AFTA/CGI in
support for its exoplanet campaign. Given that polarization filters may be
deemed an important design feature for the WFIRST coronagraph, this could
be a fortuitous alignment of circumstances. 

We note that for the more massive (such as HD 39091b) and younger giant planets
that may be WFIRST/CGI targets, their atmospheric temperatures may be high enough that
the thermal component of the fluxes at the longer wavelengths in the WFIRST/CGI optical range
may dominate the reflected component (\S\ref{heritage}; \S\ref{mass}).  This will mute the polarization fraction 
at these wavelengths, as well as the overall phase dependence of the corresponding 
planet/star flux ratios.  When measured, this effect, a function of wavelength and identifiable in
the phase dependence of both the intensity and the polarization fraction, could  
therefore be a useful diagnostic of both the planet's thermal evolution and atmospheric 
temperature.


\section{WFIRST/CGI Synergies with Other High-Contrast Platforms}
\label{synergy}

However, as discussed in \S\ref{introduction}, high-contrast imaging of
giant exoplanets in the infrared is finally emerging to complement other methods 
of exoplanet discovery and characterization (Oppenheimer \& Hinkley 2009). 
Moreover, super-Neptunes around M dwarfs might soon be within reach. Collectively, 
these direct-imaging instruments will constitute significant improvements over 
{\it Spitzer} and what is currently possible in the mid-IR. For instance, 
GPI and SPHERE will have greater planet/star sensitivities ($\sim$10$^{-6}$ $-$ 10$^{-7}$)
and smaller IWAs near $\sim$0.1$^{{\prime\prime}}$, but are optimized in the H and K bands
($\sim$1.6 $\mu$m and $\sim$2.2 $\mu$m). ELTs (such as E-ELT/EPICS and TMT/PFI) have the potential to achieve contrast
ratios of $\sim$10$^{-8}$ in the $H$ band and IWAs in the 0.02$^{{\prime\prime}}$ range.
The H band has water and CH$_4$ features, but such limited spectral coverage will put a
premium on access to other bands.

However, there are as yet no optical data for planets such as HR 8799bcde, 
$\beta$-Pictoris, GQ Lup b, 2MASS 1207b, or GJ 504b.  Expected planet/star 
contrast ratios in the optical for such planets range from 10$^{-8}$ to $10^{-6}$, 
well within the reach of WFIRST/CGI. More curiously, other than 
WFIRST/CGI, there are no credible high-contrast campaigns or capabilities 
being planned or designed for exoplanet characterization in the optical.
As Figure \ref{fig:1} demonstrates, the potential for WFIRST/CGI to fill this
gap is tremendous.  With their wide-angle separation, massive planets/brown dwarfs such as 
HR 8799bcde and their ilk will be low-lying fruit for WFIRST/CGI, yet are
examples of planets and planetary systems for which optical data can be
tightly constraining.  Being self-luminous, massive exoplanet optical/infrared colors will determine temperatures.
For the highest temperature targets, such as the HR 8799 planets with T$_{\rm eff} \sim 1000$ K,
the Na D doublet at $\sim$0.589 $\mu$m and the K I resonance doublet at $\sim$0.77 $\mu$m
should be easily identified with the WFIRST imager (by measuring the adjacent 
continuum and the lines, i.e. in and out of these absorption features), 
and the low-resolution integral-field spectrograph should clearly measure 
and trace the wide K I doublet.  At lower T$_{\rm eff}$s, the water feature 
at $\sim$0.94 \mic and the various methane features should easily be 
identifiable. In fact, comparing the optical spectra and photometric fluxes for both 
reflected-light and self-luminous targets would provide a unique window on 
the differences between giant planets over the very wide range of atmospheric thermal
and compositional classes anticipated.
  
Moreover, an interesting synergy can be achieved by combining measurements
using both WFIRST/CGI and JWST. WFIRST can provide data in the optical, while JWST will provide
data both in the near- and the mid-infrared.  There is a strong flux peak in the M band
from $\sim$4.0 to 5.5 $\mu$m to which JWST/NIRCam is sensitive.  At an IWA of
$\sim$0.7$^{\prime\prime}$, the coronagraph on NIRCam is expected to have a
planet/star contrast sensitivity near or below $\sim$10$^{-6}$ at $\sim$4.5 $\mu$m 
and will cover the important spectral region from $\sim$3.5 $\mu$m to 
$\sim$5.0 $\mu$m where there are CO, water, and CO$_2$
features in giant exoplanet spectra (Burrows, Sudarsky, \& Lunine 2003).
At smaller angles, its performance is likely to be significantly worse, but could
be $\sim$10$^{-5}$, with a stiff dependence on angular separation. The performance
of the NIRCam/JWST coronagraph at five microns will depend upon JWST's Airy
pattern and phase stability, and these have yet to be determined.  Furthermore, the long-wavelength
(5$-$28 \mic) spectrometer on JWST, MIRI, should be able to achieve planet-star contrasts from $10^{-4}$ to $10^{-6}$
in four filters centered around four wavelengths (10.65, 11.4, 15.5 and 23 $\mu$m)
with spectral resolutions ($\frac{\lambda}{\Delta\lambda}$) of $\sim$100.  The most sensitive contrast will be achieved
only at large IWAs near $\sim$1$^{{\prime\prime}}$. Again, putative 
performance curves for WFIRST/CGI, JWST, and GPI are depicted on Figure \ref{fig:1}.
Therefore, it is possible that 1) JWST could characterize a nearby, wide-separation exoplanet
at 4$-$5 $\mu$m and from $\sim$10 \mic to $\sim$23 \mic, 2) WFIRST/CGI could 
do so in the optical, and 3) the combined dataset could constrain planet properties 
much better than either alone.  There are only a few known RV planets with large enough 
greatest elongations (\S\ref{known}), but both GPI/SPHERE (in the H band) and WFIRST are likely to find more. So,
the potential of combined WFIRST and JWST data for exoplanet characterization is significant.
Since such objects are much brighter in the $M$ band and at longer wavelengths, 
this is all the more compelling for the young, and massive (perhaps brown dwarf) targets
or discoveries.

\section{Comments on Indices and Retrieval}
\label{para}

\begin{figure} [h!]
 \begin{center}
    \includegraphics[width=0.35\textwidth,angle=-90]{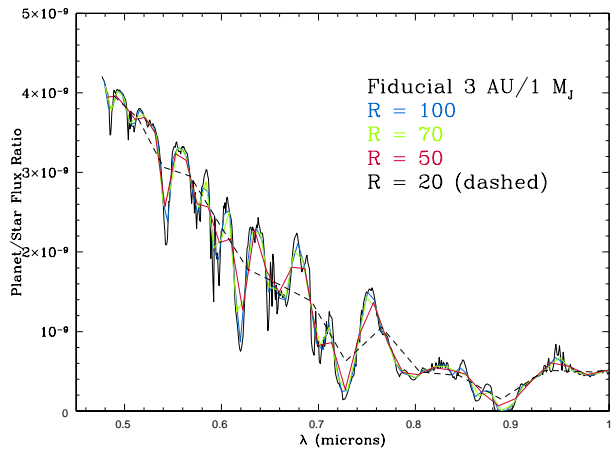}
    \vspace{-10pt}
  \end{center}
  \caption{\small A comparison of the planet/star reflection spectrum for a model
1-M$_{\rm J}$/5 Gyr giant exoplanet at 3 AU from a G2 V star, as a function of 
spectral resolution ($R = \lambda/\Delta\lambda$) from 100 to 20.  
}
    \vspace{+10pt}
    \label{rat_3AU}
\end{figure}

In advance of freezing the WFIRST coronagraph design, one may ask the question:
What spectral resolution in the optical ($R = \lambda/\Delta\lambda$) might be necessary to extract
useful information about a giant planet's atmosphere?  Below what $R$ is a spectrum nearly useless?
Figure \ref{rat_3AU} depicts the planet/star flux ratio versus wavelength at various
$R$s for a representative 1.0 M$_{\rm J}$ planet at 3.0 AU from a G2V star.
A baseline model at $R \sim 700$ is shown for reference.
One can see that the methane and water features for the $R = 100$ and $R = 70$ models are easily decernible,
but that below these $R$s features are washing out.  At $R = 20$, aside for the general slope of the curve
(which can be used to derive gross properties about the underlying cloud and possible chromophore), 
critical compositional information is quite lost.  

\begin{figure} [h!]
 \begin{center}
    \includegraphics[width=0.35\textwidth,angle=-90]{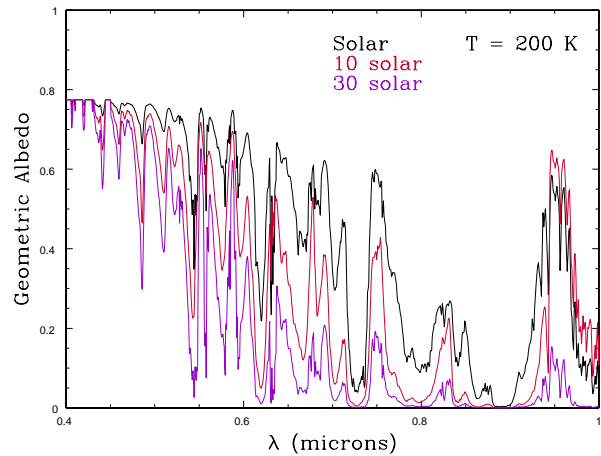}
    \vspace{-10pt}
  \end{center}
  \caption{\small A study of planet geometric albedo spectra for homogeneous
model giant exoplanet atmospheres at $T = 200$ K as a function of atmospheric
metallicity (1, 10, and 30$\times$ solar).  These models are  
cloud-free and use opacities at a pressure of 0.5 atmospheres. Gas-phase opacities
are not strong functions of pressure, but are strong functions of temperature and metallicity.  
}
    \label{alb_metal}
\end{figure}

\begin{figure} [h!]
 \begin{center}
    \includegraphics[width=0.35\textwidth,angle=-90]{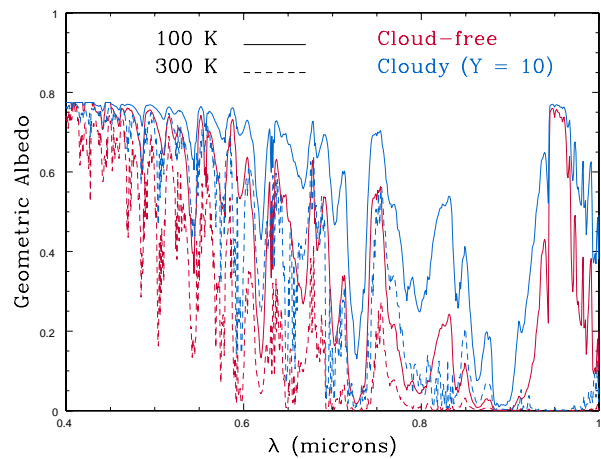}
    \vspace{-10pt}
  \end{center}
  \caption{\small A set of model geometric albedo spectra, demonstrating the generally
strong dependence of reflection albedos upon temperature or the presence of clouds.
A metallicity of 10$\times$solar and a pressure of 0.5 atmospheres are assumed.  
The parameter $Y$ here is the ratio of the cloud scattering opacity to the molecular Rayleigh 
scattering opacity and values of $Y = 0$ and $Y = 10$ are included for comparison.  
This ratio is roughly constant for homogeneously-distributed water or ammonia clouds 
consisting of small particles, but the use of the $Y$ parameter here is meant merely 
to demonstrate the general albedo spectrum behavior with cloud opacity. (We note that 
the scattering albedo of water droplets in the optical is very close to one.) Also shown are 
models with $T = 100$ K and $T = 300$ K to depict the general dependence of the albedo 
spectrum upon atmospheric temperature. The higher the temperature, the lower the albedo, 
partcularly longward of $\sim$0.65 \mic and around $\sim$0.95 \mic 
}
    \label{alb_y}
\end{figure}

Each methane feature of Figure \ref{fig:albedo_seq} is formed at a different altitude and pressure.  Therefore,
changing the methane abundance changes the strength of each feature differently. Without clouds,
this provides a straightforward way to approximate abundances $-$ various indices
(colors) calculated from the ratios of the depths of different methane spectral features 
(around $\sim$0.62 \mic, $\sim$0.74 \mic, $\sim$0.81 \mic, and $\sim$0.89 \mic) and adjacent 
continua can be calibrated to yield an average methane abundance from observations.  
Figure \ref{alb_metal} demonstrates the general dependence of the geometric albedo upon 
metallicity of a homogeneous atmosphere.  Though these models are cloud-free, this figure 
suggests that higher metallicities result in lower albedos, particularly longward of $\sim$0.6 \mic.
This figure also suggests that various flux ratios, particularly between $\sim$0.5 $-$ 0.6 \mic and/or 
$\sim$0.82 and $\sim$0.94 \mic, can be metallicity indices. Note that the higher metallicity models on this plot
seem to recapitulate the behavior of the albedo spectra at longer wavelengths of Uranus and Neptune 
(Figure \ref{jupiter_plot}) vis \`a vis those of Jupiter and Saturn. 

Figure \ref{alb_y} communicates the rough dependence
of the geometric albedo upon both atmospheric temperature and cloud thickness.  
For these example models, a metallicity of 10$\times$solar and a pressure of 0.5 atmospheres are assumed.
We note that gas-phase scattering albedos ($\omega$) are not strong functions of pressure, 
but as Figure \ref{alb_y} clearly indicates the geometric albedo of a giant exoplanet can be a stiff function
of its atmospheric temperature, particularly longward of $\sim$0.65 \mic.
Figure \ref{alb_y} also attempts to communicate the general influence of clouds on
reflectively.  The presence of clouds requires that multiple indices be used simultaneously to
break the approximate degeneracy that will result from the effect of clouds on planet/star 
flux ratios. In this case (perhaps generic), a single index or color measurement is 
less useful, since some bands can be saturated and the effects of clouds are {\it a priori} 
not known.  Hence, multiple indices are suggested for simultaneous retrieval of metallicity, atmospheric 
temperature, and cloud character. The latter could involve cloud top pressure and 
modal particle sizes, but could also involve cloud type.  Since higher temperature atmospheres
will also be thermal emitters longward of $\sim$0.8 \mic, a combination of indices that involves
the $\sim$0.9$-$1.0 \mic, $\sim$0.8$-$0.85 \mic, $\sim$0.65$-$0.7 \mic, and $\le 0.6$ \mic broad 
bands, and/or indices constructed from fluxes in and out of the methane features to 
include the associated continua, should allow one simultaneously to break various approximate 
temperature, metallicity, and cloud degeneracies.  In short, the key is multiple ``color" indices, 
strategically placed.  As articulated in \S\ref{formal}, \S\ref{kepler}, \S\ref{eccen}, 
and \S\ref{polarization}, phase and polarization data would further
improve the scientific return of a focussed campaign of exoplanet remote sensing by WFIRST.

\section{Summary and Conclusions}
\label{summary}

Here, we summarize the salient conclusions we have reached
in this quick study concerning the scientific potential of
high-contrast giant exoplanet measurement in the optical, as 
it pertains to the WFIRST-AFTA coronagraph initiative.  
A more detailed multi-parameter exploration is certainly necessary 
to flesh out the full range of potential signatures and diagnostics of 
giant exoplanet atmospheres using optical coronagraphy. However, 
despite its brevity, this quick study highlights 1) the variety
expected both for wide-separation giant planet atmospheres,
2) their discriminating planet/star ratio spectra, and 3) the rich potential 
for atmospheric characterization afforded by optical coronagraphy with WFIRST/CGI's
planned specifications.  In bulletized form, we find that:

\begin{itemize}

{\bf 
\item Albedos of giant exoplanets can be quite large and are strongly
wavelength- and orbital-distance-dependent.  Shortward of $\sim$1.0 \mic, 
most giant exoplanets reflect incident light to a larger degree in the 
$B$ and the $V$ bands than in the $R$ or $I$ bands.

\item Because the methane abundances and the presence and properties of clouds are functions
of temperature, the albedo spectrum is a non-monotonic function of planet mass and age,
orbital distance, stellar type, and planet elemental composition.

\item In the WFIRST/CGI band between $\sim$0.4 \mic and $\sim$1.0 \mic, 
the flux ratios for orbital distances from $\sim$1 AU to $\sim$10 AU
and for planet masses from 0.5 \mj to 10 \mj can vary from $\sim$10$^{-7}$ to $\sim$10$^{-10}$.  

\item The optical has prominent and diagnostic methane features near
$\sim$0.62 \mic, $\sim$0.74 \mic, $\sim$0.81 \mic, and $\sim$0.89 \mic;
ammonia features at $\sim$0.65 $\mu$m and $\sim$0.79 $\mu$m;
and a broad water band at $\sim$0.94 $\mu$m, all of which vary in strength
with orbital distance, elemental abundances, and planet mass and age.

\item The planet/star flux ratio does not follow an inverse-square law with orbital distance,
and the deviations from such behavior are functions of wavelength and composition.

\item Generally (but not universally), the optical albedo increases significantly with increasing orbital distance
from $\sim$1.0 AU to 15 AU.  With the onset of water clouds, the optical albedo rises
and the onset and thickening of reflective ammonia clouds at still larger orbital
distances results in the highest albedos at most optical and near-infrared wavelengths.

\item Since older planets have clouds with higher column depths,
giant planet albedos are expected to increase with age. Conversely, since higher
mass giants have higher surface gravities, which result in clouds
with lower column depths, at shorter wavelengths higher-mass giant exoplanets
are expected to have slightly smaller albedos and planet/star flux ratios.
This flux trend with planet mass is opposite to that expected at the longer
wavelengths (still shortward of the $\sim$1.0 \mic) accessible to WFIRST/CGI.

\item The effects of clouds on the optical and infrared can anti-correlate,
with the optical fluxes increasing and the near infrared fluxes decreasing
with increasing cloud depth.

\item All else being equal, the higher the metallicity or the higher the 
atmospheric temperature, particularly for cloud-free models,
the lower the geometric albedo.  Clouds mute this effect,
but its general operation makes the 0.94-\mic water band and 
the $\sim$0.65$-$0.8 \mic fluxes, in particular, useful functions 
of metallicity and diagnostics of atmospheric temperature.

\item Theory suggests that water clouds form around a G2V star exterior to a
distance near 1.5 AU, whereas ammonia clouds form around such a star exterior
to a distance near 4.5 AU.  Jupiter and Saturn are consistent with these expectations, but WFIRST/CGI data
as a function of orbital distance will allow us to explore this fundamental expectation
concerning cloud formation in planetary atmospheres for a much larger planet
sample and beyond the solar system.

\item Small cloud particle sizes ($\sim$1 $\mu$m) produce higher planet/star
flux ratios than large particle sizes ($\sim$100 $\mu$m) at most optical and
near-IR wavelengths.

\item Planet/star flux ratios are strong functions of orbital phase, inclination,
and eccentricity. At optical wavelengths, highly-inclined circular orbits, are 
three to four times brighter near full phase than near greatest elongation.

\item Giant exoplanets in elliptical orbits can undergo major atmospheric compositional
changes, which may have significant effects on their light curves.  Additionally,
elliptical orbits generally introduce a pronounced offset between the time of the light
curve peak and the time of full planetary phase.

\item Thermal emission, not reflection, could dominate longward of $\sim$0.7$-$0.8 \mic
for massive and/or young giant exoplanets.  This will increase the associated flux, 
change the slope of the continuum at these longer wavelengths, still within WFIRST/CGI's reach, 
and could mute the phase dependence of the corresponding emission in 
a fashion diagnostic of mass and age.

\item Whether there is a phase dependence and a day/night contrast in this thermal component
will depend upon the depth of the radiative/convective transition, itself
dependent upon the orbital distance and/or stellar flux level.
A shallow transition (expected for large separations) will diminish the 
day/night contrast of thermal flux and the corresponding phase light curve variation, 
while a deep transition (expected for closer planets) will create a 
day/night contrast at thermal wavelengths.  Hence, for more massive planets, 
the phase curve at longer wavelengths is a signature of the mechanism of day/night heat 
redistribution, either by zonal atmospheric flows or by deep convection.
Such measurements for many targets could reveal the orbital distance 
of this important transition for a given primary star type.

\item Chromophores might suppress the flux shortward of $\sim$0.6 \mic.  If present,
as they are in Jupiter and Saturn but are not in Uranus and Neptune, the associated
change in the slope of the albedo function and flux alone will reveal their presence.
Hence, correlations of the red/blue colors of WFIRST/CGI targets with stellar 
type and metallicity, planet mass, and/or orbital distance could teach us about
the associated chemistry of not only giant exoplanets, but of the solar-system giants as well.

\item Planetary polarization at a given wavelength is likely to be a more
diagnostic function of phase than the total intensity light curve in probing
cloud and haze properties and positions in gas giant atmospheres.

\item Polarization for gas molecules and for typical
cloud and aerosol particles are distinctly different, making
full linear Stokes polarization a valuable tool for determining the presence
and nature of aerosol particles.

\item Peak exoplanet polarizations can be large, ranging from $\sim$10\% to $\sim$90\%,
and vary diagnostically with wavelength.

\item At shorter optical wavelengths,
polarization generally contains information about the upper planet atmosphere,
while at longer wavelengths clouds deeper in the atmosphere are probed.
Hence, polarization color indices (for example, comparing $R$ and $B$, or $R$ and $I$-band)
can reveal the presence, thickness, and optical depth of a Rayleigh scattering layer.

\item There is a phase angle at which the polarized flux (as opposed to the 
polarized fraction) is a maximum.  For conservative Rayleigh scattering, this 
angle is $\sim$65$^{\circ}$, but it is generally larger for haze particles.
Hence, this angle alone can be used as a diagnostic of the presence of aerosols.

\item Since polarization is a relative measure, and, hence, perhaps more 
intrinsically accurate, the degree of polarization in both spectral bands
and the continuum depends strongly upon atmospheric composition and structure.

\item When combining geometric albedo and polarization data as a function of wavelength
simultaneously (for a known orbital phase), one should be able to break various approximate 
degeneracies introduced by the presence of aerosols. 

\item Tropospheric and stratospheric cloud models might be distinguished by the line-to-continuum
polarization ratios.  Though the fractional polarization can be large in strong absorption bands,
while the flux itself is small (due to absorption), the continuum does not suffer from this effect.
Hence, both flux and polarization ratio indices between the absorption bands and the adjacent continuum
are useful diagnostics of scattering cloud depth, the presence of aerosols, particle size and make-up,
and the relative contribution of Rayleigh scattering.

\item The sharpness of the phase variation in the polarization increases 
markedly for larger particles or shorter wavelengths, and this can 
be used to directly constrain particle size.

\item The robustness of the qualitative single-scattering polarization in the 
context even of multiple scattering suggests that spherical droplets could imprint 
rainbow and/or glory effects as a function of wavelength.

\item At longer wavelengths, and for more massive and/or younger giant planets,
the fact that the flux could be thermal erases the polarization and its orbital 
phase dependence. Thus, the polarization contrast between the short and 
long WFIRST/CGI wavelength bins, and its variation with phase, should be 
diagnostic functions of the temperature of the planet's atmosphere and of planet mass and/or age.

}

\end{itemize}

\acknowledgments

The author would like to thank David Sudarsky, Nikku Madhu, Wes Traub, Ivan Hubeny, Jonathan Lunine,
Mark Marley, Glenn Schnieder, Brian MacIntosh, Jim Breckinridge, Hans-Martin Schmid for collaborations, conversations,
and insights and acknowledges JPL for support during the execution of this project.
He gratefully acknowledges NASA's Exoplanet Exploration Program for supporting this study on behalf of the
WFIRST/AFTA Science Definition Team and the Exo-S and Exo-C Science and Technology Definition Teams.

\begin{deluxetable}{cccccccc}
\tablewidth{18.5cm}
\tablecaption{Model Data for Specific RV Gas Giants\label{data.specific}}
\tablehead{
\colhead{Planet} & \colhead{Max. separation ($^{\prime\prime})$} & \colhead{star}& \colhead{a (AU)} & \colhead{d (pc)}
& \colhead{P (yrs.)} & \colhead{M$_p$$\sin(i)$ (\mj)} & \colhead{$log_{10}$ $g$ (cgs)}}  
\startdata

55 Cnc d          & 0.44 & G8V & 5.9 & 13.4 & 14.7 & 4.05 & 4.30 \\  
47 UMa c          & 0.28 & G0V& 3.73& 13.3 & 7.10 & 0.76  & 3.48 \\  
Gl 777A b         & 0.23& G6V& 3.65& 15.9 & 7.15 & 1.15   & 3.48 \\  
$\upsilon$ And d  & 0.19 & F8V& 2.50& 13.5 & 3.47 & 4.61  & 4.30 \\  
HD 39091b         & 0.16& G1IV& 3.34& 20.6 & 5.70 & 10.3  & 4.48 \\  
47 UMa b          & 0.16 & G0V& 2.09& 13.3 & 2.98 & 2.54  & 3.78 \\  
$\gamma$ Cephei b & 0.15 & K2V& 1.8& 11.8 & 2.5 & 1.25    & 3.60 \\  
14 Her b          & 0.15 & K0V& 2.5 & 17   & 4.51 & 3.3   & 3.90 \\  

\enddata
\end{deluxetable}

\begin{deluxetable}{cccc}
\tablewidth{16.0cm}
\tablecaption{Geometric Albedo (A$_g$), Polarization Fraction Maximum (P$_{\rm max}$), and $\theta_{\rm max}$ for Vector Rayleigh Scattering
off a Spherical Planet, as a Function of Scattering Albedo ($\omega$) \label{vec_ray}}
\tablehead{\colhead{$\omega$} & \colhead{A$_g$} & \colhead{P$_{\rm max}$}& \colhead{$\theta_{\rm max}$}}
\startdata
      0.100   & 0.0198  &    0.9787  &   90.0\\
      0.200   & 0.0420  &    0.9513  &   89.0\\
      0.300   & 0.0672  &    0.9216  &   89.0\\
      0.400   & 0.0963  &    0.8884  &   90.0\\
      0.500   & 0.1304  &    0.8498  &   90.0\\
      0.600   & 0.1716  &    0.8038  &   90.0\\
      0.700   & 0.2234  &    0.7472  &   90.0\\
      0.800   & 0.2928  &    0.6750  &   90.0\\
      0.900   & 0.3999  &    0.5733  &   91.0\\
      0.980   & 0.5857  &    0.4320  &   92.0\\
      0.990   & 0.6403  &    0.3995  &   93.0\\
      0.995   & 0.6823  &    0.3772  &   93.0\\  
      1.000   & 0.7977  &    0.3263  &   95.0\\
\enddata
\end{deluxetable}


\begin{thebibliography}{}

\def\apj{Astrophys.~J.}
\def\apjl{Astrophys.~J.~Lett.}
\def\aj{Astron.~J.\ }
\def\mnras{Mon.~Not.~R.~Astron.~Soc.}
\def\ann{Annu.~Rev.~Astron.~Astrophys.}
\def\araa{Annu.~Rev.~Astron.~Astrophys.}
\def\apjs{Astrophys.~J.~Suppl.~Ser.}
\def\pasp{Publ.~Astron.~Soc.~Pac.}
\def\prl{Phys. Rev. Lett.}
\def\aap{{Astron. Astrophys.}}
\def\aa{{Astron. Astrophys.}}
\def\reference{\bibitem}
\def\citep{\cite}
\def\bf{, }

\renewcommand\refname{\vskip -0.3 in}



\bibitem{SPHERE} Beuzit, J.-L. et al. SPHERE: a planet finder instrument for the VLT.
in Ground-based and Airborne Instrumentation for Astronomy II. Edited by McLean, Ian S.; Casali, Mark M.
{\it Proc. SPIE} 7014:701418-701418-12 (2008)




\bibitem{barman_hr8799b} Barman, T., Macintosh, B., Konopacky, Q.M., \& Marois, C.
Clouds and Chemistry in the Atmosphere of Extrasolar Planet HR8799b. \apj, 733, article id. 65, 18 pp. (2011)



\bibitem[Buenzli \& Schmid(2009)]{buenzli} Buenzli, E. \& Schmid, H.M. 2009, \aap, 504, 259




\bibitem[Burrows \etal(1997)]{Burrows97} Burrows, A., Marley, M.,
Hubbard, W. B., Lunine, J. I., Guillot, T., Saumon, D., Freedman, R.,
Sudarsky, D., \& Sharp, C. 1997, ``A Non-Gray Theory of Extrasolar Giant Planets and Brown Dwarfs," \apj, 491, 856

\bibitem[Burrows \& Sharp(1999)]{bursharp99} Burrows, A. \& Sharp, C.M. 1999,
``Chemical Equilibrium Abundances in Brown Dwarf and Extrasolar Giant Planet Atmospheres,'' \apj, 512, 843.





\bibitem[Burrows, Sudarsky, \& Lunine(2003)]{burrows2003} Burrows, A.,
Sudarsky, D., and Lunine, J.I. 2003, ``Beyond the T Dwarfs: Theoretical Spectra, Colors,
and Detectability of the Coolest Brown Dwarfs," \apj, 596, 587

 
\bibitem[Burrows, Sudarsky, \& Hubeny(2004)]{bur2004} Burrows, A.,
Sudarsky, D., \& Hubeny, I. 2004, ``Spectra and Diagnostics for the Direct Detection of Wide-Separation Extrasolar Giant Planets," \apj, 609, 407



\bibitem{burrows_2005} Burrows, A. A theoretical look at the direct detection of giant
planets outside the Solar System. Nature, 433, 261-268 (2005)













\bibitem[Cooper et al.(2003)]{cooper} Cooper, C.S., Sudarsky, D.,
Milsom, J.A., Lunine, J.I., \& Burrows, A. 2003, ``Modeling the Formation of Clouds in Brown Dwarf Atmospheres," \apj, 586, 1320




\bibitem{deming_JWST} Deming, D. et al. Discovery and Characterization of Transiting Super Earths
Using an All-Sky Transit Survey and Follow-up by the James Webb Space Telescope. PASP, 121, 952-967 (2009)


\bibitem[Deirmendjian(1964)]{deir} Deirmendjian, D. 1964, Appl. Opt., 3, 187

\bibitem[Deirmendjian(1969)]{deir69} Deirmendjian, D. 1969, Electromagnetic scattering on spherical polydispersions, 
New York, NY (USA): Elsevier Scientific Publishing



 


\bibitem[Fluri \& Berdyugina(2010)]{fluri} Fluri, D.M. \& Berdyugina, S.V. 2010, \aap, 512, A59 pp. 1-12











\bibitem[Hansen \& Hovenier(1971)]{hansen1971} Hansen, J.E. \& Hovenier, J.W. 1971, J. Quant. Spectrosc. Radiat. Transfer, 11, 809

\bibitem[Hansen \& Hovenier(1974)]{hansenh} Hansen, J.E. \& Hovenier, J.W. 1974, J. Atmos. Sci., 31, 1137

\bibitem[Hansen \& Travis(1974)]{hansent} Hansen, J.E. \& Travis, L.D. 1974, Space Sci. Rev., 16, 527






 



\bibitem[Karkoschka(1994)]{Karkoschka94} Karkoschka, E. 1994, Icarus,
111, 174

\bibitem[Karkoschka(1998)]{Karkoschka98} Karkoschka, E. 1998, Icarus,
133, 134


\bibitem[Khare et al.(1984)]{khare84} Khare, B.N. et al. 1984, Icarus, 60, 127








\bibitem[Kurucz(1994)]{Kurucz94} Kurucz, R. 1994, ``Solar abundance model atmospheres for 0,1,2,4,8 km/s," {\it Kurucz CD-ROM
No. 19}, (Cambridge: Smithsonian Astrophysical Observatory)

\bibitem{betapic} Lagrange, A.M. et al. A probable giant planet imaged in the $\beta$ Pictoris disk. VLT/NaCo deep L$^{\prime}$-band imaging.
\aap, 493, L21-L25 (2009)







\bibitem{GPI} Macintosh, B. et al. The Gemini Planet Imager: from science to design to construction.
in Adaptive Optics Systems. Edited by Hubin, Norbert; Max, Claire E.; Wizinowich, Peter L. {\it Proc. SPIE} 7015:701518-701518-13 (2008)

\bibitem{analytic_albedos} Madhusudhan, N., \& Burrows, A.
Analytic Models for Albedos, Phase Curves, and Polarization of Reflected Light from Exoplanets.
\apj, 747, 25-40 (2011)

\bibitem{hr8799_madhu} Madhusudhan, N., Burrows, A., \& Currie, T.
Model Atmospheres for Massive Gas Giants with Thick Clouds:  Application to the HR 8799 Planets.
\apj, 737, 34-48 (2011)



\bibitem{marley_1999} Marley, M.S., Gelino, C., Stephens, D., Lunine, J.I., \& Freedman, R.
Reflected Spectra and Albedos of Extrasolar Giant Planets. I. Clear and Cloudy Atmospheres.
\apj, 513, 879-893 (1999)



\bibitem{marois} Marois, C. et al. Direct Imaging of Multiple Planets Orbiting the Star HR 8799.
Science, 322, 1348-1352 (2008)

\bibitem{marois_e} Marois, C., Zuckerman, B., Konopacky, Q.M., Macintosh, B., \& Barman, T.
Images of a fourth planet orbiting HR 8799. Nature, 468, 1080-1083 (2010)





\bibitem[Oppenheimer \& Hinkley(2009)]{oppen} Oppenheimer, B.R. \& Hinkley, S. 2009,
``High-Contrast Observations in Optical and Infrared Astronomy," Ann. Rev. Astron. Astrophys., 47, 253
\bibitem[Patten et al.(2006)]{patten} Patten, B. et al. 2006, ``Spitzer IRAC Photometry of M, L, and T



\bibitem[Pope et al.(1992)]{Pope92} Pope, S. K., Tomasko, M. G.,
Williams, M. S., Perry, M. L., Doose, L. R., \& Smith, P. H.
1992, Icarus, 100, 203

\bibitem[Rages(1997)]{Rages97} Rages K.A., Galileo Imaging Team 1997,
AAS DPS, 29, 19.17

\bibitem[Rages, Pollack, \& Smith(1983)]{Rages83} Rages K.A., Pollack, J.B, \& Smith, P.H. 1983, J. Geophys. Res., 88, 8721




\bibitem[Sagan \& Khare(1979)]{sagan} Sagan, C. \& Khare, B.N. 1979, Nature, 277, 102-107


\bibitem[Schmid et al.(2006)]{schmid} Schmid, H.M. et al. 2006, ``Search and Investigation of 
Extra-solar Planets with Polarimetry," in Proceedings on Direct Imaging of Exoplanets: 
Science \& Techniques, IAU Colloquium No. 200, 2005), doi:10,1017/S1743921306009252




\bibitem[Skrutskie et al.(2010)]{skrut} Skrutskie, M.F. et al. ``The Large Binocular Telescope mid-infrared camera (LMIRcam): final design and status,"
in Ground-based and Airborne Instrumentation for Astronomy III. Edited by McLean, Ian S.; 
Ramsay, Suzanne K.; Takami, Hideki. Proceedings of the SPIE, Volume 7735, article id. 77353H, 11 pp. (2010).





\bibitem[Sharp \& Burrows(2007)]{sharp07} Sharp, C.M. \& Burrows, A. 2007,
``Atomic and Molecular Opacities for Brown Dwarf and Giant Planet Atmospheres," \apjs, 168, 140

\bibitem[Smith(1986)]{Smith86} Smith, P. H. 1986, Icarus, 65, 264

\bibitem[Smith \& Tomasko(1984)]{smith_tomasko} Smith, P.H. \& Tomasko, M.G. (1984), Icarus, 58, 35

\bibitem{spergel} Spergel, D.N., Gehrels, N. et al. Wide-Field InfraRed Survey Telescope $-$ Astrophysics
Focused Telescope Assets WFIRST-AFTA Final Report. http://arxiv.org/abs/1305.5422 (astroph/1305.5422) (2013)

\bibitem[Stam, Hovenier, \& Waters(2004)]{stam} Stam, D.M., Hovenier, J.W., \& Waters, L.B.F.M. 2004, \aap, 28, 663-672

\bibitem[Sudarsky \etal(2000)]{Sudarsky00} Sudarsky, D., Burrows, A.,
\& Pinto, P. 2000, ``Albedo and Reflection Spectra of Extrasolar Giant Planets," \apj, 538, 885

\bibitem[Sudarsky, Burrows, \& Hubeny(2003)]{sud03}
Sudarsky, D., Burrows, A., and Hubeny, I. 2003, ``Theoretical Spectra and Atmospheres of Extrasolar Giant Planets," \apj, 588, 1121

\bibitem[Sudarsky et al.(2005)]{sud05}
Sudarsky, D., Burrows, A., Hubeny, I., \& Li, A. 2005,
``Phase Functions and Light Curves of Wide-Separation Extrasolar Giant Planets," \apj, 627, 520

\bibitem{suzuki} Suzuki, R. et al. Performance characterization of the HiCIAO instrument for the Subaru Telescope.
Proceedings of the SPIE, 7735, article id. 773530, 13 pp. (2010)




\bibitem[Tomasko \& Smith(1982)]{tomasko82} Tomasko, M.G. \& Smith, P.H. 1982, Icarus, 51, 65




\bibitem[West et al.(1983)]{west83} West, R.A., Hart, H., Simmons, K.E. et al. 1983, J. Geophys. Res., 88, 8699

\bibitem[West et al.(1986)]{West86} West, R. A., Strobel, D. F,
\& Tomasko, M. G. 1986, Icarus 65, 161

\bibitem[West \& Smith(1991)]{west91} West, R.A. \& Smith, P.H. 1991, Icarus, 90, 330
 

\bibitem[Wiktorowicz(2009)]{Wiktorowicz:09} Wiktorowicz, S.J. 2009, \apj, 696, 1116



\end{thebibliography}
\end{document}